\documentclass[12pt]{article}

\usepackage{graphicx}
\usepackage{amsfonts}
\usepackage{amssymb}

\def\gtwid{\mathrel{\raise.3ex\hbox{$>$\kern-.75em\lower1ex\hbox{$\sim$}}}}
\def\ltwid{\mathrel{\raise.3ex\hbox{$<$\kern-.75em\lower1ex\hbox{$\sim$}}}}
\def\square{\kern1pt\vbox{\hrule height 1.2pt\hbox{\vrule width 1.2pt\hskip 3pt
   \vbox{\vskip 6pt}\hskip 3pt\vrule width 0.6pt}\hrule height 0.6pt}\kern1pt}

\begin{document}

\begin{titlepage}

\begin{flushright}
UFIFT-QG-16-02 , CCTP-2016-05 \\
CCQCN-2016-142 , ITCP-IPP 2016/06
\end{flushright}

\vskip 2cm

\begin{center}
{\bf The Effect of Features on the Functional Form of the Scalar Power Spectrum}
\end{center}

\vskip 1cm

\begin{center}
D. J. Brooker$^{1*}$, N. C. Tsamis$^{2\star}$ and R. P. Woodard$^{1\dagger}$
\end{center}

\vskip .5cm

\begin{center}
\it{$^{1}$ Department of Physics, University of Florida,\\
Gainesville, FL 32611, UNITED STATES}
\end{center}

\begin{center}
\it{$^{2}$ Institute of Theoretical Physics \& Computational Physics, \\
Department of Physics, University of Crete, \\
GR-710 03 Heraklion, HELLAS}
\end{center}

\vspace{1cm}

\begin{center}
ABSTRACT
\end{center}
We study how the scalar power spectrum of single-scalar inflation depends 
functionally on models with features which have been proposed to explain
anomalies in the data. We exploit a new formalism based on evolving the 
norm-squared of the scalar mode functions, rather than the mode functions 
themselves.

\begin{flushleft}
\end{flushleft}

\vskip .5cm

\begin{flushleft}
$^{*}$ e-mail: djbrooker@ufl.edu \\
$^{\star}$ e-mail: tsamis@physics.uoc.gr \\
$^{\dagger}$ e-mail: woodard@phys.ufl.edu
\end{flushleft}

\end{titlepage}

\section{Introduction}

It would be difficult to over-emphasize the importance of primordial
perturbations predicted by inflation \cite{Starobinsky:1979ty,
Mukhanov:1981xt}. These are not only the first detectable quantum 
gravitational phenomena ever identified \cite{Woodard:2009ns,
Ashoorioon:2012kh,Krauss:2013pha}, they also provide the initial
conditions for structure formation in the standard model of 
cosmology \cite{Mukhanov:1990me,Liddle:1993fq,Dodelson:2003ft}.
It is therefore frustrating, and even somewhat embarrassing, that 
exact results are unavailable for any realistic model of inflation, 
even at tree order, and even when one knows the model.

For definiteness, let us assume that inflation is described by general
relativity minimally coupled to some scalar potential model,
\begin{equation}
\mathcal{L} = \frac{R \sqrt{-g}}{16 \pi G} - \frac12 \partial_{\mu}
\varphi \partial_{\nu} \varphi g^{\mu\nu} \sqrt{-g} - V(\varphi)
\sqrt{-g} \; . \label{Lagrangian}
\end{equation}
At tree order the tensor and scalar power spectra, $\Delta^2_{h}(k)$
and $\Delta^2_{\mathcal{R}}(k)$, are known in terms of the constant 
amplitudes approached by their mode functions, $u(t,k)$ and $v(t,k)$, 
after the first horizon crossing time $t_k$ \cite{Woodard:2014jba},
\begin{eqnarray}
\Delta^2_{h}(k) & = & \frac{k^3}{2\pi^2} \times 32\pi G \times
2 \times \Bigl\vert u(t,k)\Bigr\vert^2_{t \gg t_k} 
\approx \frac{16 G H^2(t_k)}{\pi} \; , \label{Dh} \\
\Delta^2_{\mathcal{R}}(k) & = & \frac{k^3}{2 \pi^2} \times 4 \pi G
\times \Bigl\vert v(t,k)\Bigr\vert^2_{t \gg t_k} \approx 
\frac{G H^2(t_k)}{\pi \epsilon(t_k)} \; , \label{DR}
\end{eqnarray}
where $k = H(t_k) a(t_k)$. The problem is that we do not have exact 
solutions for the mode functions for realistic models of inflation, so 
their asymptotic values must be computed numerically. The functional 
forms standing to the right of expressions (\ref{Dh}) and (\ref{DR}) 
are only the leading WKB approximations.

We use the Hubble representation \cite{Liddle:1994dx}, which is based
on knowing the scale factor $a(t)$, and hence also the Hubble parameter
$H(t)$ and the first slow roll parameter $\epsilon(t)$,\footnote{
The more familiar potential representation can be reached by reconstructing 
the scalar and its potential from the background Einstein equations
\cite{Tsamis:1997rk,Saini:1999ba,Capozziello:2005mj,Woodard:2006nt,Guo:2006ab},
\begin{eqnarray}
\varphi_0(t) = \varphi_0(t_i) \pm \int_{t_i}^{t} \!\! dt' H(t')
\sqrt{ \frac{\epsilon(t')}{4 \pi G}} \quad \Longleftrightarrow \quad t(\varphi)
\qquad , \qquad 
V(\varphi) = \frac{[3 \!-\! \epsilon(t)] H^2(t)}{8 \pi G} \Biggl\vert_{
t=t(\varphi)} \; . \nonumber
\end{eqnarray}}
\begin{equation}
ds^2 = -dt^2 + a^2(t) d\vec{x} \!\cdot\! d\vec{x} \quad \Longrightarrow \quad
H(t) \equiv \frac{\dot{a}}{a} \quad , \quad \epsilon(t) \equiv -
\frac{\dot{H}}{H^2} \; . \label{geometry}
\end{equation}
The evolution equation, Wronskian and asymptotically early form of the
tensor mode functions $u(t,k)$ are \cite{Woodard:2014jba},
\begin{equation}
\ddot{u} + 3 H \dot{u} + \frac{k^2}{a^2} u = 0 \; , \; u \dot{u}^* \!-\! 
\dot{u} u^* = \frac{i}{a^3} \; , \; u(t,k) \longrightarrow 
\frac{\exp[-ik \int_{t_i}^t \frac{dt'}{a(t')}]}{\sqrt{2 k a^2(t)}} \; .
\label{ueqns}
\end{equation}
The analogous relations for the scalar mode functions $v(t,k)$ are 
\cite{Woodard:2014jba},
\begin{equation}
\ddot{v} + \Bigl(3 H + \frac{\dot{\epsilon}}{\epsilon}\Bigr) \dot{v} + 
\frac{k^2}{a^2} v = 0 \; , \; v \dot{v}^* \!- \dot{v} v^* = \frac{i}{\epsilon a^3} 
\; , \; v(t,k) \longrightarrow \frac{\exp[-ik \int_{t_i}^t \frac{dt'}{a(t')}]}{
\sqrt{2 k \epsilon(t) a^2(t)}}  \; . \label{veqns}
\end{equation}
Relations (\ref{Dh}-\ref{DR}) imply that $u(t,k)$ and $v(t,k)$ approach 
constants after $k^2/a^2$ becomes negligible. But the only exact solutions 
are for $\epsilon(t) = \epsilon_0$,
\begin{eqnarray}
u_0(t,k;\epsilon_0) & \equiv & 
\sqrt{\frac{\pi}{4 (1 \!-\! \epsilon_0) H(t) a^3(t)}} \; 
H^{(1)}_{\nu_0}\Bigl( \frac{k}{(1 \!-\! \epsilon_0) H(t) a(t)}\Bigr) 
 \; , \qquad  \label{constepsu} \\
v_0(t,k;\epsilon_0) & \equiv & 
\frac{u_0(t,k;\epsilon_0)}{\sqrt{\epsilon_0}} \qquad , \qquad \nu_0 
\equiv \frac12 \Bigl(\frac{3\!-\!\epsilon_0}{1\!-\!\epsilon_0} \Bigr) \; .
\label{constepsv}
\end{eqnarray} 
It is by now quite clear that no constant value of $\epsilon(t)$ is consistent
with the data \cite{Ade:2015lrj}.

Although numerical methods must employed, that does not mean the best
strategy is to evolve $u(t,k)$ and $v(t,k)$, or that one should abandon 
the goal of deriving good analytic approximations which are valid for 
an arbitrary inflationary geometry. We have developed a formalism for
evolving the norm-squares, $M(t,k) \equiv \vert u(t,k)\vert^2$ and 
$\mathcal{N}(t,k) \equiv \vert v(t,k)\vert^2$, which avoids the need to 
keep track of the rapidly changing and irrelevant phases \cite{Romania:2012tb}. 
We have also shown that factoring out the instantaneously constant $\epsilon$ 
solution $u_0(t,k;\epsilon(t))$, and making a judicious choice of variables, 
gives rise to a wonderfully accurate analytic approximation for the tensor 
power spectrum \cite{Brooker:2015iya,Brooker:2016imi}. The purpose of this 
paper is to do the same for the scalar power spectrum.

A major issue we wish to resolve is which of the two possible approaches
gives the best analytic approximation for the scalar power spectrum:
\begin{enumerate}
\item{Exploiting the functional relation between the scalar and tensor 
mode functions \cite{Tsamis:2003zs,Romania:2012tb}; or}
\item{Factoring out the instantaneously constant $\epsilon(t)$ solution as
we did for the tensor power spectrum \cite{Brooker:2015iya}.}
\end{enumerate}
In section 2 we describe these two strategies. Section 3 examines them
numerically, and Section 4 discusses the results.

\section{Different Strategies for Computing $\Delta^2_{\mathcal{R}}(k)$}

Previous work has shown how relations (\ref{ueqns}-\ref{veqns}) can be
used to derive equations for the norm-squares $M(t,k) \equiv \vert u(t,k)
\vert^2$ and $\mathcal{N}(t,k) \equiv \vert v(t,k)\vert^2$
\cite{Romania:2012tb},
\begin{eqnarray}
\ddot{M} + 3 H \dot{M} + \frac{2 k^2}{a^2} M = \frac{(\dot{M}^2 \!+\! 
\frac1{a^6})}{2 M} & , & M(t,k) \longrightarrow \frac1{2 k a^2(t)} 
\; , \label{Meqn} \\
\ddot{\mathcal{N}} + \Bigl(3H \!+\! \frac{\dot{\epsilon}}{\epsilon} \Bigr) 
\dot{\mathcal{N}} \!+\! \frac{2 k^2}{a^2} \, \mathcal{N}  = 
\frac{( \dot{\mathcal{N}}^2 \!+\! \frac1{\epsilon^2 a^6})}{2 \mathcal{N}} 
& , & \mathcal{N}(t,k) \longrightarrow \frac1{2 k \epsilon(t) a^2(t)} 
\; . \qquad \label{Neqn}
\end{eqnarray}
These equations are much more efficient than (\ref{ueqns}-\ref{veqns}) for
computing the power spectra because there is no need to keep track of the
irrelevant phase. (The improvement is roughly quadratic 
\cite{Brooker:2015iya}.) The purpose of this section is to explain the two 
possible strategies for further improvement in computing $\mathcal{N}(t,k)$: 
either by transforming $M(t,k)$ or by factoring out the instantantenously 
constant $\epsilon$ solution. 

\subsection{Transforming from Tensor to Scalar}

It is important to emphasize that we are not thinking about the numerical 
values of the power spectra for some specific model but rather about how 
the power spectra depend {\it functionally} on an arbitrary inflationary 
expansion history $a(t)$. We use square brackets to denote this functional 
dependence,
\begin{equation}
M[a](t,k) \qquad , \qquad \mathcal{N}[a](t,k) \; .
\end{equation}
It is easy to check that the tensor mode relations (\ref{ueqns}) become
those of the scalar (\ref{veqns}) under a simultaneous redefinition of 
``time'' and of the expansion history \cite{Tsamis:2003zs,Romania:2012tb},
\begin{equation}
d\widetilde{t} \equiv \sqrt{\epsilon(t)} \, dt \qquad , \qquad
\widetilde{a}(\widetilde{t}) \equiv \sqrt{\epsilon(t)} \, a(t) \; .
\label{trans}
\end{equation}
The functional relation between $M$ and $\mathcal{N}$ is therefore,
\begin{equation}
\mathcal{N}[a](t,k) = M[\widetilde{a}](\widetilde{t},k) \; . \label{MtoN}
\end{equation}
We have a good functional approximation for how the tensor power 
spectrum depends on a general inflationary expansion history 
\cite{Brooker:2015iya}. So one might try computing $\mathcal{N}[a](t,k)$
by applying (\ref{MtoN}) to this approximate form.

In practice one would never actually solve (\ref{trans}) for 
$\widetilde{a}(\widetilde{t})$. The two geometrical quantities needed for
our expression of $M[\widetilde{a}](\widetilde{t},k)$ are the Hubble
parameter $\widetilde{H}(\widetilde{t})$ and the first slow roll parameter
$\widetilde{\epsilon}(\widetilde{t})$. It is better to express them as
functions of the untransformed time $t$, using the geometrical parameters
of the original expansion history. For example, the transformed Hubble 
parameter is,
\begin{equation}
\widetilde{H}(\widetilde{t}) \equiv \frac1{\sqrt{\epsilon(t)}} \frac{d}{dt}
\ln\Bigl[ \sqrt{\epsilon(t)} \, a(t)\Bigr] = \frac{H(t)}{\sqrt{\epsilon(t)}}
\Bigl[1 \!+\! \frac{\dot{\epsilon}(t)}{2 \epsilon(t) H(t)} \Bigr] \; .
\label{Htilde}
\end{equation}
The corresponding transformed first slow roll parameter is,
\begin{equation}
\widetilde{\epsilon}(\widetilde{t}) \equiv \frac1{\sqrt{\epsilon}} 
\frac{d}{dt} \Bigl[ \frac{ \sqrt{\epsilon}}{H \!+\! 
\frac{\dot{\epsilon}}{2 \epsilon}} \Bigr] = \frac{ \epsilon \!+\! 
\frac{\dot{\epsilon}}{2 \epsilon H} \!+\! 3 (\frac{\dot{\epsilon}}{2 
\epsilon H})^2 \!-\! \frac{\ddot{\epsilon}}{2 \epsilon H^2}}{[1 \!+\!
\frac{\dot{\epsilon}}{2 \epsilon H}]^2} \; . \label{epstilde}
\end{equation}
Our form for $M[a](t,k)$ also requires the first and second derivatives of
the transformed slow roll parameter, which can be computed by substituting 
expression (\ref{epstilde}) in,
\begin{equation}
\frac{d \widetilde{\epsilon}(d \widetilde{t})}{d \widetilde{t}} =
\frac1{\sqrt{\epsilon(t)}} \frac{d\widetilde{\epsilon}(\widetilde{t})}{dt} 
\qquad , \qquad 
\frac{d^2 \widetilde{\epsilon}(\widetilde{t})}{d \widetilde{t}^2} =
\frac1{\sqrt{\epsilon(t)}} \frac{d}{dt} \Biggl[
\frac1{\sqrt{\epsilon(t)}} \frac{d\widetilde{\epsilon}(\widetilde{t})}{dt}
\Biggr] \; .
\end{equation}
These expressions obviously involve third and fourth derivatives of 
$\epsilon(t)$.

Of course relation (\ref{MtoN}) is exact, so applying it to an {\it exact}
result for $M[a](t,k)$ must recover $\mathcal{N}[a](t,k)$. However, what 
we actually have is a good {\it approximate} form for $M[a](t,k)$ 
\cite{Brooker:2015iya}, and our approximations for the $a(t)$ expansion 
history might not be so good for $\widetilde{a}(\widetilde{t})$. In
particular, there are plausible models (we will study one in section 3) 
for which $\epsilon(t)$ is small but its first derivative can become large 
enough to make the factors of $[1 + \frac{\dot{\epsilon}}{2 \epsilon H}]$ 
in expressions (\ref{Htilde}-\ref{epstilde}) pass through zero. So one can 
see how an accurate approximation for $M[a](t,k)$ might lead to very 
inaccurate results for $\mathcal{N}[a](t,k)$ using relation (\ref{MtoN}).

\subsection{Factoring Out $\mathcal{N}_0(t,k) \equiv \vert
v_0(t,k,\epsilon(t)) \vert^2$}

A different strategy is to parallel the derivation for $M[a](t,k)$ 
\cite{Brooker:2015iya}. Our technique in that case was to factor out
the instantaneously constant $\epsilon(t)$ solution,
\begin{equation}
M(t,k) \equiv M_0(t,k) \!\times\! \Delta M(t,k) \qquad , \qquad
M_0(t,k) \equiv \Bigl\vert u_0\Bigl(t,k;\epsilon(t)\Bigr) \Bigr\vert^2
\; , \label{Mstep1} 
\end{equation}
where $u_0(t,k;\epsilon_0)$ is the known solution (\ref{constepsu}) 
for $\epsilon(t) = \epsilon_0$. Substituting (\ref{Mstep1}) in 
(\ref{Meqn}) and dividing by $M(t,k)$ leads to an evolution equation 
for the residual $\Delta M(t,k)$,
\begin{eqnarray}
\lefteqn{\frac{\Delta \ddot{M}}{\Delta M} - \frac12 \Bigl( \frac{ 
\Delta \dot{M}}{\Delta M}\Bigr)^2 + \Bigl(3 H \!+\! \frac{\dot{M}_0}{M_0} 
\Bigr) \frac{\Delta \dot{M}}{\Delta M} - \frac1{2 a^6 M_0^2} 
\Bigl( \frac1{\Delta M^2} \!-\! 1\Bigr) } \nonumber \\
& & \hspace{4cm} = \frac1{2 a^6 M_0^2} - \frac{2 k^2}{a^2} -
\Bigl[ \frac{\ddot{M}_0}{M_0} \!-\! \frac12 \Bigl( \frac{\dot{M}_0}{M_0}
\Bigr)^2 \!+\! 3H \frac{\dot{M}_0}{M_0}\Bigr] \; . \qquad \label{Mstep2}
\end{eqnarray}
One can show that the right hand side of (\ref{Mstep2}) vanishes for
constant $\epsilon(t)$ \cite{Brooker:2015iya}. The next step is to 
convert from co-moving time $t$ to the number of e-foldings from the
beginning of inflation, $n \equiv \ln[a(t)/a_i]$. Denoting derivatives
with respect to $n$ by a prime and dividing by $H^2$ gives,
\begin{eqnarray}
\lefteqn{\frac{\Delta M''}{\Delta M} - \frac12 \Bigl( \frac{ 
\Delta M'}{\Delta M}\Bigr)^2 + \Bigl(3 \!-\! \epsilon \!+\! 
\frac{M'_0}{M_0} \Bigr) \frac{\Delta M'}{\Delta M} - \frac1{2 H^2 a^6 M_0^2}
 \Bigl( \frac1{\Delta M^2} \!-\! 1\Bigr) } \nonumber \\
& & \hspace{2.5cm} = \frac1{2 H^2 a^6 M_0^2} - \frac{2 k^2}{H^2 a^2} -
\Bigl[ \frac{M''_0}{M_0} \!-\! \frac12 \Bigl( \frac{M'_0}{M_0}\Bigr)^2 
\!+\! (3 \!-\! \epsilon) \frac{M'_0}{M_0}\Bigr] \; . \qquad \label{Mstep3}
\end{eqnarray}
The many ratios in (\ref{Mstep3}) suggests a change of dependent variable 
to $\Delta M(t,k) = \exp[-\frac12 h(n,k)]$, which implies,
\begin{eqnarray}
\lefteqn{h'' - \frac14 (h')^2 + \Bigl(3 \!-\! \epsilon \!+\! \frac{M'_0}{M_0} 
\Bigr) h' + \frac1{H^2 a^6 M_0^2} \Bigl( e^h \!-\! 1\Bigr) 
= S(n,k)} \nonumber \\
& & \hspace{2.5cm} \equiv \frac{4 k^2}{H^2 a^2} - \frac1{H^2 a^6 M_0^2} + 2
\Bigl[ \frac{M''_0}{M_0} \!-\! \frac12 \Bigl( \frac{M'_0}{M_0}\Bigr)^2 
\!+\! (3 \!-\! \epsilon) \frac{M'_0}{M_0}\Bigr] \; . \qquad \label{Mstep4}
\end{eqnarray}

Relation (\ref{Mstep4}) is exact, but it can be easily converted to a 
successive approximation scheme that gives $h(n,k)$ for a {\it general} 
expansion history. It is useful to first take note of the relation which
exists between the ``frequency'' and ``friction'' terms in (\ref{Mstep4}),
\begin{equation}
\frac1{H(t) a^3(t) M_0(t,k)} \equiv \omega(n,k) \qquad \Longrightarrow \qquad
3 - \epsilon + \frac{M_0'}{M_0} = -\frac{\omega'}{\omega} \; . \label{keysimp}
\end{equation}
Now move the terms nonlinear in $h(n,k)$ from the left side of (\ref{Mstep4})
to the right side,
\begin{equation}
h'' - \frac{\omega'}{\omega} h' + \omega^2 h = S(n,k) + \frac14 {h'}^2 + 
\omega^2 \Bigl[1 \!+\! h \!-\! e^h\Bigr] \; . \label{Mstep5}
\end{equation}
The left hand side of (\ref{Mstep5}) is a linear differential operator acting
on $h(n,k)$, whose retarded Green's function can be written down for a general
expansion history \cite{Brooker:2015iya},
\begin{equation}
G(n;n') = \frac{\theta(n \!-\! n')}{\omega(n',k)} \, \sin\Biggl[ \int_{n'}^{n}
\!\!\!\! dn'' \omega(n'',k)\Biggr] \; . \label{Gfunct}
\end{equation}
Because $M_0(t,k)$ recovers the full asymptotic early time form 
(\ref{Meqn}), the initial conditions for $h(n,k)$ are $h(0,k) = 0 = h'(0,k)$. 
We can use the Green's function to express the solution to equation 
(\ref{Mstep5}) as a series $h = h_1 + h_2 + \dots$ in powers of the source. 
The first two terms are,
\begin{eqnarray}
h_1(n,k) & = & \int_0^{n} \!\! dn' G(n;n') S(n',k) \; , \label{h1} \\
h_2(n,k) & = & \int_0^{n} \!\! dn' G(n;n') \Biggl[ \frac14 {h_1'}^2(n',k)
- \frac12 \omega^2(n',k) h_1^2(n',k)\Biggr] \; . \label{h2}
\end{eqnarray}

Note that the series expansion gives $h(n,k)$ as a functional of the 
expansion history, for a general inflationary $a(t)$. Our previous study
\cite{Brooker:2015iya} has shown that this expansion is about twice as 
convergent as the generalized slow roll expansion \cite{Stewart:2001cd,
Dvorkin:2009ne}. Equation (\ref{Mstep5}) also provides a powerful way of
understanding the behavior of $h(n,k)$ as a damped, driven oscillator. The 
frequency term $\omega(n,k)$ in (\ref{keysimp}) is huge until just a few 
e-foldings before horizon crossing and then decays to zero exponentially,
while the friction term is always of order one. 

The review we have just presented of the tensor analysis is very relevant
because the instantaneously constant $\epsilon(t)$ solution for the scalar
is closely related to its tensor cousin,
\begin{equation}
\mathcal{N}_0(t,k) \equiv \Bigl\vert v_0\Bigl(t,k;\epsilon(t)\Bigr)
\Bigr\vert^2 = \frac{M_0(t,k)}{\epsilon(t)} \; . \label{Nstep1}
\end{equation}
If one factors this term out,
\begin{equation}
\mathcal{N}(t,k) \equiv \frac{M_0(t,k)}{\epsilon(t)} \!\times\! 
\Delta \mathcal{N}(t,k) \; ,
\end{equation}
and then divides (\ref{Neqn}) by $N(t,k)$, the resulting equation for
$\Delta \mathcal{N}(t,k)$ is almost identical to (\ref{Mstep2}),
\begin{eqnarray}
\lefteqn{\frac{\Delta \ddot{\mathcal{N}}}{\Delta \mathcal{N}} - \frac12 
\Bigl( \frac{ \Delta \dot{\mathcal{N}}}{\Delta \mathcal{N}}\Bigr)^2 + 
\Bigl(3 H \!+\! \frac{\dot{M}_0}{M_0} \Bigr) \frac{\Delta \dot{\mathcal{N}}}{
\Delta \mathcal{N}} - \frac1{2 a^6 M_0^2} \Bigl( \frac1{\Delta \mathcal{N}^2} 
\!-\! 1\Bigr) } \nonumber \\
& & \hspace{0cm} = \frac1{2 a^6 M_0^2} - \frac{2 k^2}{a^2} -
\Bigl[ \frac{\ddot{M}_0}{M_0} \!-\! \frac12 \Bigl( \frac{\dot{M}_0}{M_0}
\Bigr)^2 \!+\! 3H \frac{\dot{M}_0}{M_0}\Bigr] + 
\frac{\ddot{\epsilon}}{\epsilon} - \frac12 \Bigl( 
\frac{\dot{\epsilon}}{\epsilon}\Bigr)^2 + 
3 H \frac{\dot{\epsilon}}{\epsilon} \; . \qquad \label{Nstep2}
\end{eqnarray}
This means we can immediately read off the result of changing the independent
variable to $n \equiv \ln[a(t)/a_i]$ and the dependent variable to $g(n,k) 
\equiv -2 \ln[\Delta \mathcal{N}(t,k)]$,
\begin{equation}
g'' - \frac{\omega'}{\omega} g' + \omega^2 g = S(n,k) + \Delta S(n)
+ \frac14 {g'}^2 + \omega^2 \Bigl[1 \!+\! g \!-\! e^g\Bigr] \; . \label{Nstep5}
\end{equation}
Only the extra (and $k$-independent) source $\Delta S(n)$ distinguishes 
the scalar equation (\ref{Nstep5}) from its tensor counterpart (\ref{Mstep5}),
\begin{equation}
\Delta S(n) = -2 \Biggl[ \frac{\epsilon''}{\epsilon} - \frac12 \Bigl(
\frac{\epsilon'}{\epsilon}\Bigr)^2 + (3 \!-\! \epsilon) \frac{\epsilon'}{
\epsilon} \Biggr] \; . \label{Nsource}
\end{equation}
It is useful to express $\Delta S(n) = \Delta S_1(n) + \Delta S_2(n)$ as
the sum of a total derivative and a negative-definite term,
\begin{equation}
\Delta S_1\equiv -2\partial_n \Bigl[ \partial_n \ln(\epsilon) \!+\! 3 
\ln(\epsilon) \!-\! \epsilon\Bigr] \quad , \quad 
\Delta S_2 \equiv - \Bigl[ \partial_n \ln(\epsilon)\Bigr]^2 \; . \label{DS12}
\end{equation}
Although $\Delta S_1$ is typically larger, the net impulse comes entirely
from $\Delta S_2$.

In addition to the derivation, we can also read off the scalar asymptotic analysis 
from the tensor case. At early times, before horizon crossing, the frequency, 
friction and source take the forms \cite{Brooker:2015iya},
\begin{eqnarray}
\omega(n,k) & \longrightarrow & \frac{2 k}{H a} + 
O\Bigl( \frac{H a}{k}\Bigr) \; , \label{Mfreqearly} \\
-\frac{\omega'(n,k)}{\omega(n,k)} & \longrightarrow & 1 - \epsilon +
O\Bigl( \frac{H^2 a^2}{k^2}\Bigr) \; , \label{Mfrictionearly} \\
S(n,k) & \longrightarrow & -\Bigl[ \epsilon'' + (9 \!-\! 7 \epsilon)
\epsilon'\Bigr] \Bigl( \frac{H a}{k}\Bigr)^2 + O\Bigl( 
\frac{H^4 a^4}{k^4}\Bigr) \; . \label{Msourceearly}
\end{eqnarray}
Hence derivatives of $g(n,k)$ are irrelevant with respect to the 
restoring force, and the tensor source is irrelevant with respect to the 
extra scalar part (\ref{Nsource}),
\begin{equation}
{\rm Early\ Times:} \quad g(n,k) = \Delta S(n) 
\!\times\! \Bigl( \frac{H a}{2 k}\Bigr)^2 + O\Bigl( \frac{H^4 a^4}{k^4}
\Bigr) \; . \label{gearly}
\end{equation}

At late times, after horizon crossing, the continuing evolution of
$\epsilon(t)$ causes $\mathcal{N}_0(t,k)$ to evolve slightly in a way that 
must be cancelled by continuing evolution in $g(n,k)$ in order for 
$\mathcal{N}(t,k)$ to approach a constant. This behavior can be described 
by the function $F(n,k)$ defined as \cite{Brooker:2015iya},
\begin{equation}
F(n,k) \equiv \ln\Bigl[ \frac{2 (1 \!-\! \epsilon) H a}{k}\Bigr] + 
\psi\Bigl( \frac12 \!+\! \frac1{1 \!-\! \epsilon}\Bigr) - 1 \; . 
\label{Fdef}
\end{equation}
The late time forms for the frequency, friction and source are
\cite{Brooker:2015iya},
\begin{eqnarray}
\omega(n,k) & \longrightarrow & \frac{\pi}{\Gamma(\frac12 \!+\! 
\frac1{1 - \epsilon})} \Bigl( \frac{2 k}{H a}\Bigr) \Bigl[ \frac{k}{2 
(1 \!-\! \epsilon) H a}\Bigr]^{\frac2{1-\epsilon}} + 
O\Bigl[ \Bigl( \frac{k}{H a}\Bigr)^{3 + \frac2{1-\epsilon}} \Bigr] 
\; , \label{Mfreqlate} \\
-\frac{\omega'(n,k)}{\omega(n,k)} & \longrightarrow & \frac{2 \epsilon' F}{
(1 \!-\! \epsilon)^2} + 3 - \epsilon + O\Bigl(\frac{k^2}{H^2 a^2}\Bigr) 
\; , \label{Mfrictionlate} \\
S(n,k) & \longrightarrow & \frac{4 \epsilon'' F}{(1 \!-\! \epsilon)^2} +
\frac{4 \epsilon'}{1 \!-\! \epsilon} \Bigr[ \Bigl( \frac{3 \!-\! \epsilon}{
1 \!-\! \epsilon}\Bigr) F \!+\! 1\Bigr] \nonumber \\
& & \hspace{.5cm} + \frac{4 {\epsilon'}^2}{(1 \!-\!
\epsilon)^3} \Bigl[ \frac{F^2}{1 \!-\! \epsilon} \!+\! 2 F \!-\! 1 \!+\!
\frac{\psi(\frac12 \!+\! \frac1{1 -\epsilon})}{1 \!-\! \epsilon}\Bigr]
+ O\Bigl(\frac{k^2}{H^2 a^2}\Bigr) \; . \qquad \label{Msourcelate}
\end{eqnarray}
This means the restoring force is irrelevant. Although the extra scalar 
source (\ref{Nsource}) does not grow, neither does it fall off, so it 
contributes a term to the late time behavior of $g(n,k)$, in addition to 
those already implied by the known form of $h(n,k)$ \cite{Brooker:2015iya},
\begin{eqnarray}
\lefteqn{{\rm Late\ Times:} \quad  g(n,k) = 2 \ln\Biggl[
\Bigl( \frac{a(t)}{a(t_k)}\Bigr)^{\frac{2 \epsilon(t)}{1-\epsilon(t)}}
\Bigl( \frac{H(t)}{H(t_k)}\Bigr)^{\frac{2}{1-\epsilon(t)}}
\frac{C(\epsilon(t))}{C(\epsilon(t_k)} \frac{\epsilon(t_k)}{\epsilon(t)}
\Biggr] } \nonumber \\
& & \hspace{7.5cm} - 2\ln\Bigl[ \mathcal{S}(k)\Bigr] + 
O\Bigl( \frac{k^2}{H^2 a^2}\Bigr) \; . \qquad \label{glate}
\end{eqnarray}
For small $\epsilon(t)$ this extra term is enough to give $g(n,k)$ a 
downward slope,
\begin{equation}
{\rm Late\ Times:} \quad g'(n,k) = -\frac{2 \epsilon'}{\epsilon} +
\frac{4 \epsilon' F}{(1 \!-\! \epsilon)^2} + O\Bigl( \frac{k^2}{H^2 a^2}
\Bigr) \; . \label{g'late}
\end{equation}

The functions $C(\epsilon)$ and $\mathcal{S}(k)$ which appear in 
expression (\ref{glate}) show up as well in the scalar power spectrum,
\begin{equation} 
\Delta^2_{\mathcal{R}}(k) = \frac{G H^2(t_k)}{\pi \epsilon(t_k)}
\!\times\! C\Bigl( \epsilon(t_k)\Bigr) \!\times\! \mathcal{S}(k) \; .
\label{DRform}
\end{equation}
$C(\epsilon)$ gives all local corrections (that is, evaluated at 
$t = t_k$) to the leading slow roll approximation on the right of 
(\ref{DR}),
\begin{equation}
C(\epsilon) \equiv \frac1{\pi} \Gamma^2\Bigl( \frac12 \!+\! 
\frac1{1 \!-\! \epsilon}\Bigr) \Bigl[ 2 (1 \!-\! \epsilon)
\Bigr]^{\frac2{1-\epsilon}} \; . \label{Cdef}
\end{equation}
The quantity $\mathcal{S}(k)$ represents the nonlocal part of the
power spectrum, the need for which has long been recognized 
\cite{Wang:1997cw}. It is determined by evolving $g(n,k)$ to late times 
--- either numerically or analytically --- and then comparing with 
expression (\ref{glate}). The analytic series expansion $g = g_1 + g_2 
+ \dots$, can be developed from (\ref{Nstep5}) the same way we did for
the tensor case. Its first two terms are,
\begin{eqnarray}
g_1(n,k) & = & \int_0^{n} \!\! dn' G(n;n') \Bigl[S(n',k) + \Delta S(n)
\Bigr] \; , \label{g1} \\
g_2(n,k) & = & \int_0^{n} \!\! dn' G(n;n') \Biggl[ \frac14 {g_1'}^2(n',k)
- \frac12 \omega^2(n',k) g_1^2(n',k)\Biggr] \; . \label{g2}
\end{eqnarray}
Examination of the nonlinear terms on the right hand side of equation 
(\ref{Nstep5}) shows that the linear approximation $g(n,k) \approx 
g_1(n,k)$ can only fail under two conditions:
\begin{enumerate}
\item{If $g'(n,k)$ becomes order one or larger; or}
\item{If $g(n,k)$ becomes order one or larger during the few e-folding
around horizon crossing when $\omega(n,k)$ is also of order one.}
\end{enumerate}
Because $g(n,k)$ is negligible before horizon crossing, the second 
possibility is difficult to realize, but we will see that the presence of
a feature can cause the first to occur.

\section{Comparing Analytic and Numerical Results}

The purpose of this section is to use our formalism to study models
with features. We shall mostly focus on a class of ``step'' models 
\cite{Adams:2001vc} whose parameters were fit to explain peculiarities
in WMAP data \cite{Mortonson:2009qv}. The section accordingly begins with
a review of this model. Then we contrast exact numerical simulation of
its scalar power spectrum with our linearized approximation and with an
improvement based on correcting the source. The evolution of a series 
of nine different modes is studied to develop a qualitative understanding 
of the ``ringing'' phenomenon. The section closes with a brief examination
of other models with features.

\begin{figure}[ht]
\includegraphics[width=6.0cm,height=4.8cm]{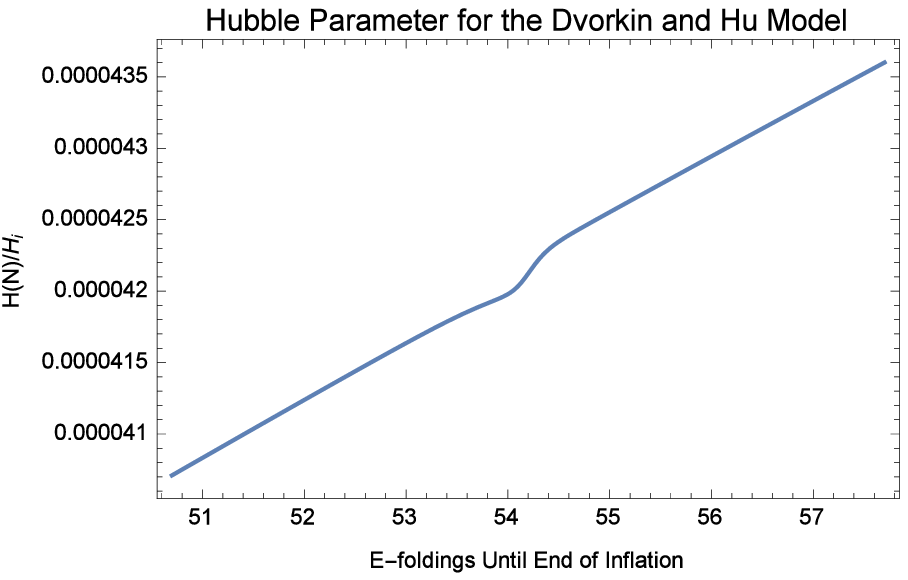}
\hspace{1cm}
\includegraphics[width=6.0cm,height=4.8cm]{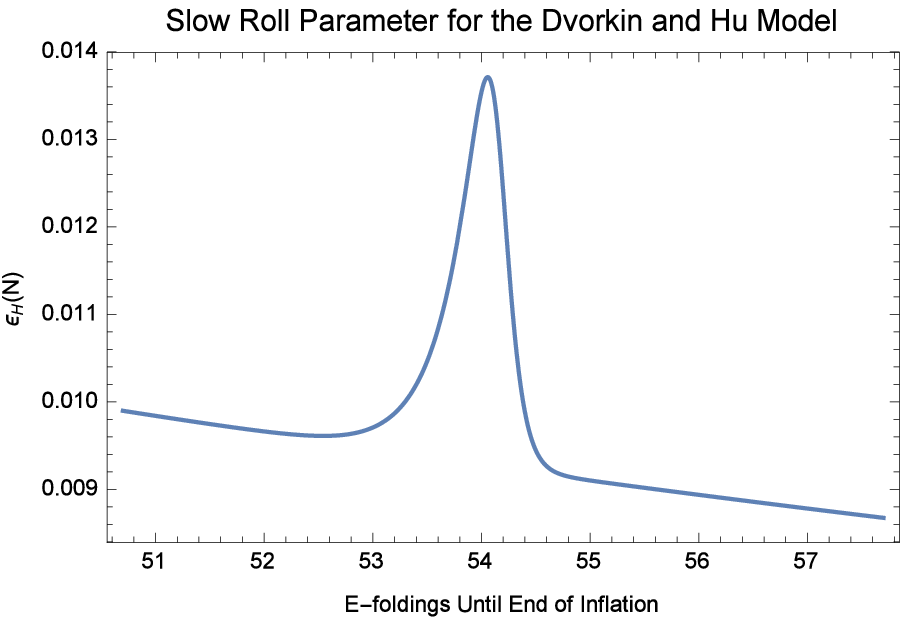}
\caption{Graphs of the Hubble parameter and the first slow roll 
parameter as functions of the number of e-foldings $N \equiv n_{\rm end}
- n$ until the end of inflation for the step model of section 3.1. Note 
that early times (large $N$) are at the right and late times (small $N$) 
are at the left.}
\label{modelA}
\end{figure}

\begin{figure}[ht]
\includegraphics[width=6.0cm,height=4.8cm]{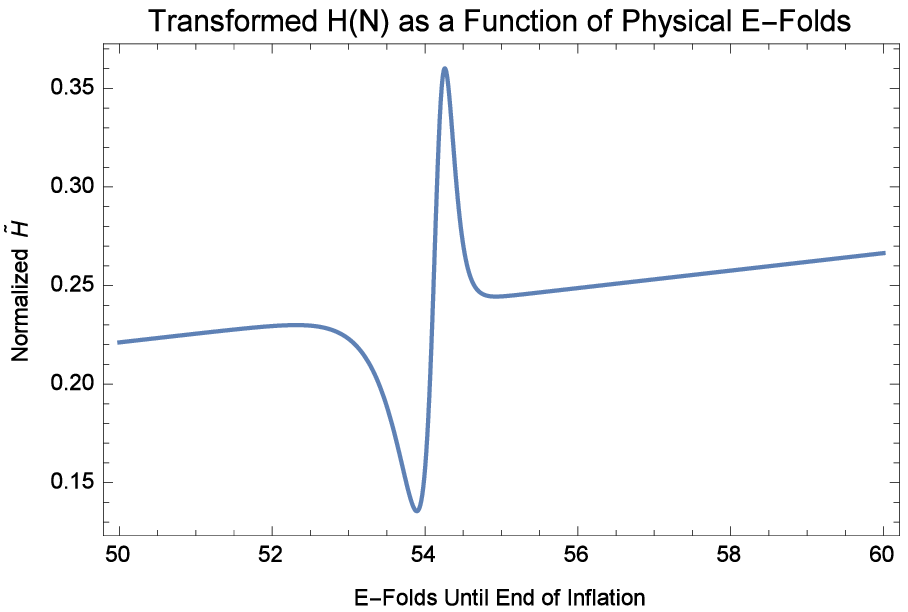}
\hspace{1cm}
\includegraphics[width=6.0cm,height=4.8cm]{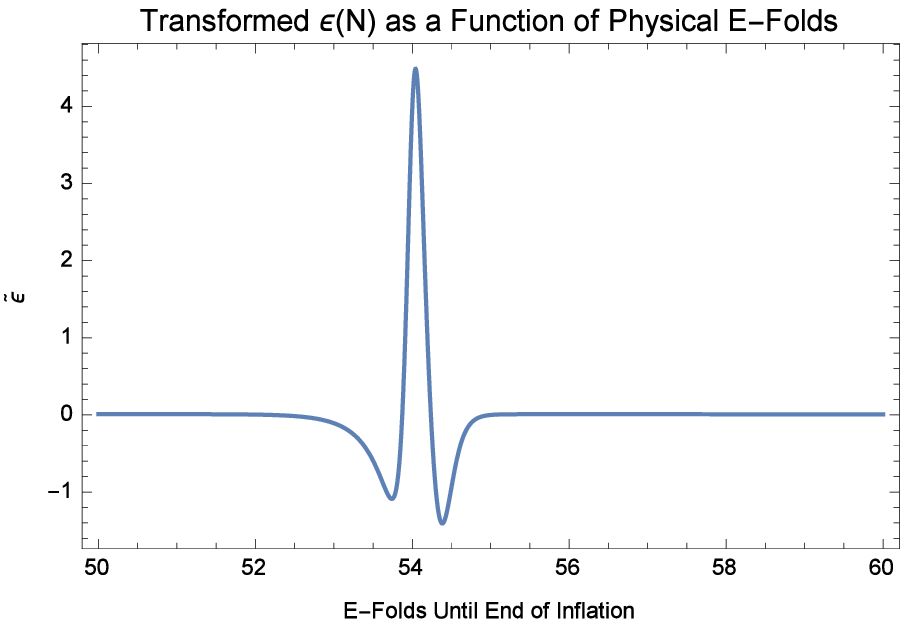}
\caption{Graphs of the Hubble parameter $\widetilde{H}(\widetilde{t})$
(\ref{Htilde}) and the transformed first slow roll parameter 
$\widetilde{\epsilon}(\widetilde{t})$ (\ref{epstilde}) for the impulse
model of section 3.4. These are the geometrical parameters which would
be used to compute the scalar power spectrum by transforming the tensor
power spectrum according to relation (\ref{MtoN}). Both graphs are 
expressed as functions of the number of (untransformed) e-foldings $N 
\equiv n_{\rm end} - n$ until the end of inflation.}
\label{modelC}
\end{figure}

\subsection{The Step Model}

In 2001 Adams, Cresswell and Easther proposed a generic model whose
potential is quadratic with a multiplicative step of variable location, 
height and width \cite{Adams:2001vc},
\begin{equation}
V(\varphi) = \frac12 m^2 \varphi^2 \Bigl[1 + c \tanh\Bigl( 
\frac{\varphi \!-\! b}{d}\Bigr) \Bigr] \; . \label{potential}
\end{equation}
Features in the range $20 \lesssim \ell \lesssim 40$ of the WMAP 3-year
data \cite{Covi:2006ci,Hamann:2007pa} were fitted to give the following
maximum likelihood values for the parameters $b$, $c$, $d$ and $m$
\cite{Mortonson:2009qv},
\begin{eqnarray}
b = \frac{14.668}{\sqrt{8 \pi G}} \qquad & , & \qquad 
c = 1.505 \!\times\! 10^{-3} \;, \label{params1} \\
d = \frac{0.02705}{\sqrt{8 \pi G}} \qquad & , & \qquad 
m = \frac{7.126 \!\times\! 10^{-6}}{\sqrt{8 \pi G}} \; . \label{params2}
\end{eqnarray}

\begin{figure}[ht]
\includegraphics[width=4cm,height=3cm]{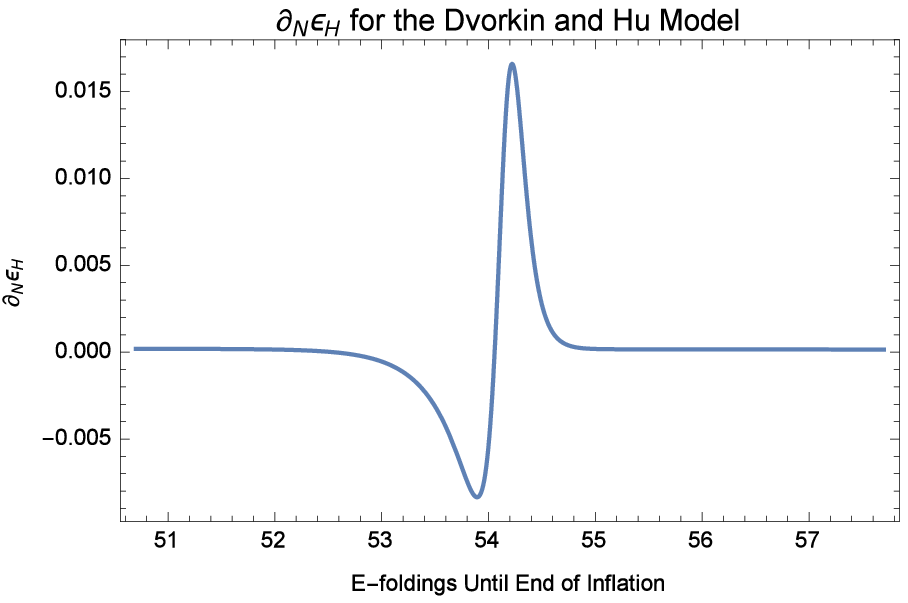}
\hspace{.6cm}
\includegraphics[width=4cm,height=3cm]{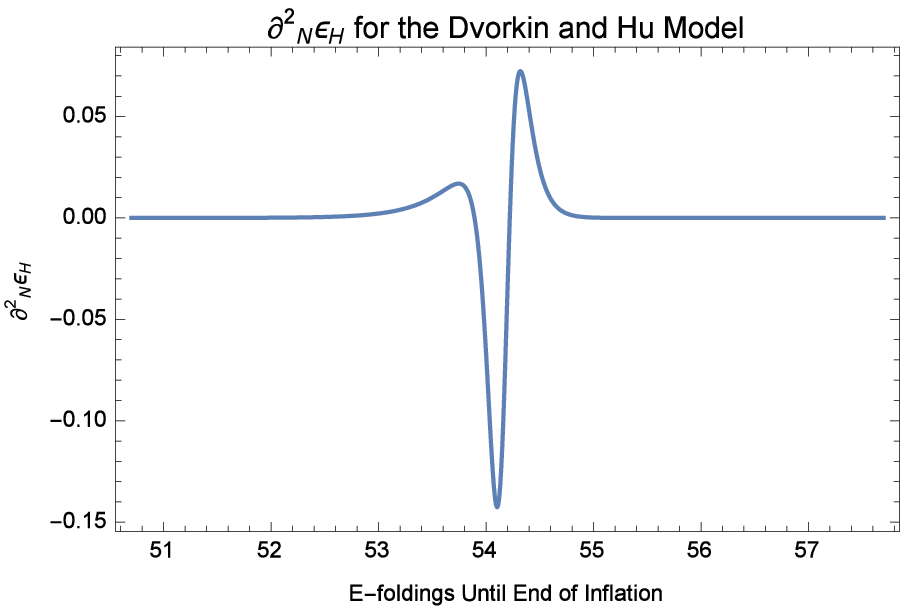}
\hspace{.6cm}
\includegraphics[width=4cm,height=3cm]{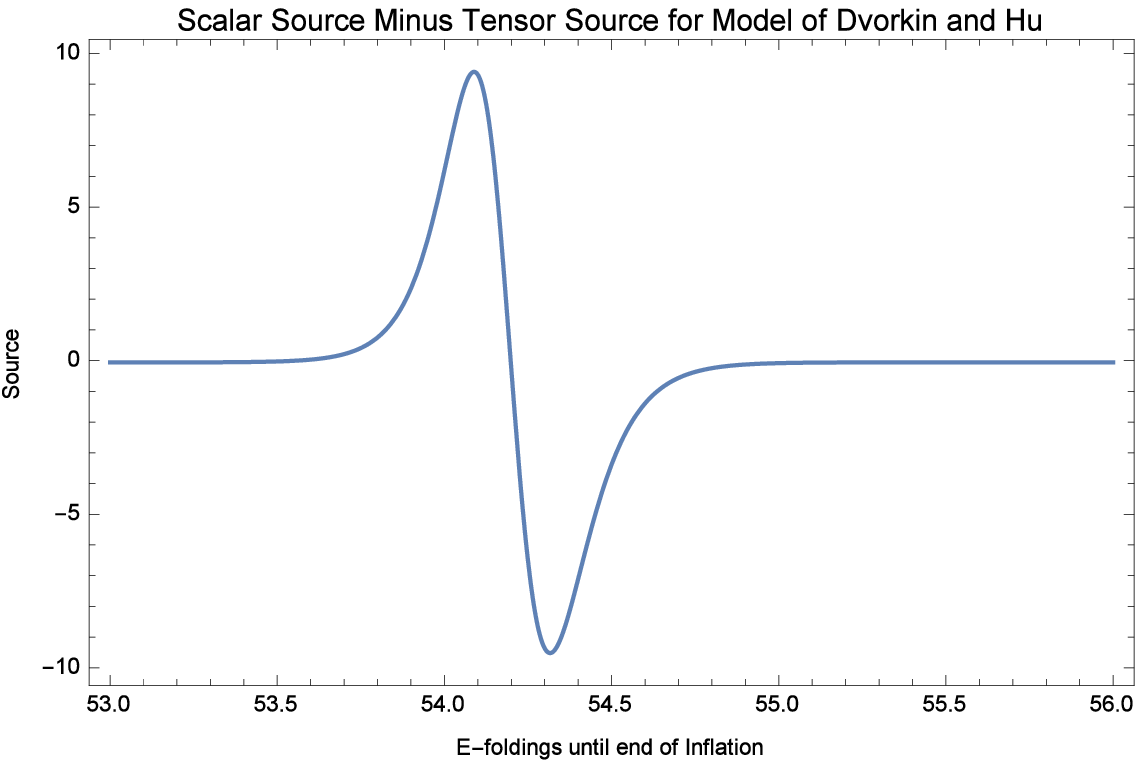}
\caption{Graphs of $\epsilon'$ (left), $\epsilon''$ (middle), and the 
extra source $\Delta S(n)$ (right) defined in equation (\ref{Nsource}) 
for the step model of section 3.1. In each case the functions are expressed 
in terms of the number of e-foldings $N \equiv n_{\rm end} - n$ until the 
end of inflation.}
\label{modelB}
\end{figure}

In numerically evolving this model it is best to convert to dimensionless
fields and potentials,
\begin{equation}
\psi(n) \equiv \sqrt{8\pi G} \, \varphi(t) \qquad , \qquad 
U(\psi) \equiv (8\pi G)^2 V(\varphi) \; . \label{dimless}
\end{equation}
When we also convert from co-moving time $t$ to the number of e-foldings
$n \equiv \ln[a(t)/a_i]$, the exact scalar equation becomes,\footnote{
One really only needs $\psi(0)$ to solve (\ref{psieqn}). The other initial 
condition can be taken as $\psi'(0) = U'(\psi(0))/U(\psi(0))$ using the
slow roll approximation, which ought to be excellent long before the 
onset of the feature.}
\begin{equation}
\psi'' + \Bigl(3 \!-\! \frac12 {\psi'}^2\Bigr) \psi' + \Bigl(3 \!-\! 
\frac12 {\psi'}^2\Bigr) \frac{U'(\psi)}{U(\psi)} = 0 \; . \label{psieqn}
\end{equation}
The Hubble parameter and first slow roll parameter follow from $\psi(n)$
through the exact relations,
\begin{equation}
8\pi G H^2 = \frac{U(\psi)}{3 \!-\! \frac12 {\psi'}^2} \qquad , \qquad
\epsilon = \frac12 {\psi'}^2 \; . \label{stepgeom}
\end{equation}
Figure~\ref{modelA} shows these as functions of the number of e-foldings 
from the end of inflation $N \equiv n_{\rm end} - n$. Note that there is
only a slight bump in the Hubble parameter, and that the first slow roll
parameter remains small even though the feature does enhance it by as
much as 50\% over the single e-folding in the range $53.7 \lesssim N 
\lesssim 54.7$. 

Figure~\ref{modelC} shows transformed geometrical parameters 
(\ref{Htilde}-\ref{epstilde}) that would be used to obtain the scalar 
power spectrum by exploiting the transformation (\ref{MtoN}) from the 
tensor power. Note that the transformed Hubble parameter changes
by almost a factor of three within a single e-folding, as the model
oscillations from normal acceleration $0 \leq \widetilde{\epsilon} < 1$ 
to super-acceleration $\widetilde{\epsilon} < 0$ and even deceleration
$1 < \widetilde{\epsilon}$. Our linearized approximation for the tensor
power spectrum is very good \cite{Brooker:2015iya}, but it breaks down
for such wild fluctuations. Thus we conclude that the strategy laid out
in section 2.2 is better.

Figure~\ref{modelB} shows the first two derivatives of $\epsilon$ with
respect to $n$, along with the extra scalar source $\Delta S(n)$ defined
in equation (\ref{Nsource}). All functions are again displayed in terms
of the number of e-foldings until the end of inflation $N \equiv n_{\rm end}
- n$. A number of things are worth noting. First, the smallness of 
$\vert \partial_n \epsilon\vert < 0.017$ and of $\vert \partial_n^2 
\epsilon \vert < 0.15$ means that the tensor source $S(n,k)$ is negligible
with respect to the scalar part $\Delta S(n)$ (\ref{Nsource}), whose 
magnitude can reach almost 10. Second, the fact that $\epsilon''$ has
almost ten times the magnitude of $\epsilon'$ means that, of the source's
two parts (\ref{DS12}), the first one $\Delta S_1(n)$ totally dominates the 
second one $\Delta S_2(n)$. This is significant because $\Delta S_1(n)$ is 
a total derivative, which contributes zero net impulse after the feature 
has passed, whereas $\Delta S_2$ contributes a net negative impulse. We 
will see in section 4 that features for which $\Delta S_1(n)$ and 
$\Delta S_2(n)$ are comparable induce a qualitatively different response 
in $g(n,k)$ and $\Delta^2_{\mathcal{R}}(k)$.

\begin{figure}[ht]
\includegraphics[width=6.5cm,height=5.2cm]{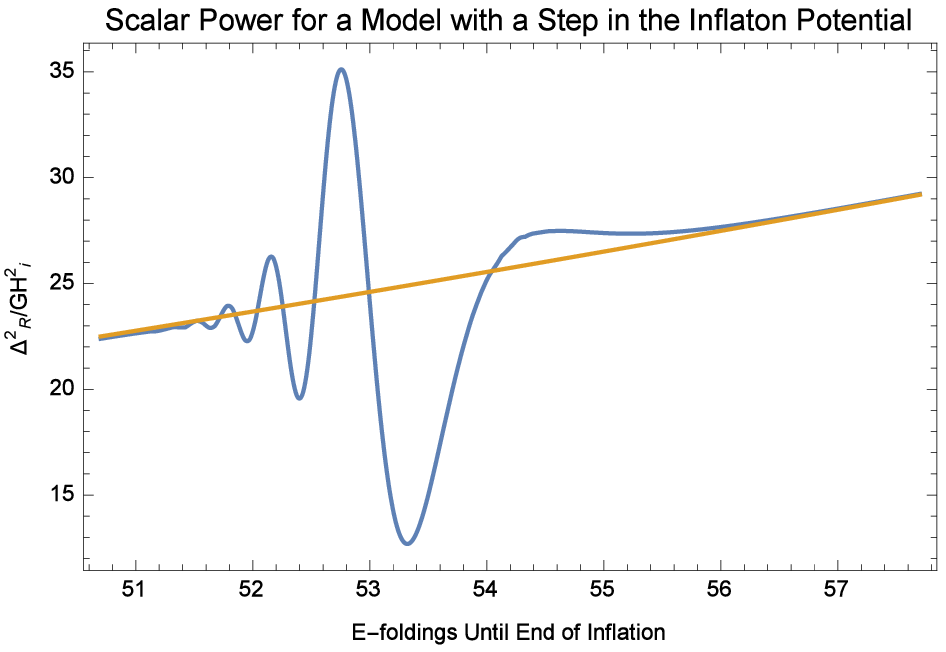}
\hspace{.5cm}
\includegraphics[width=6.5cm,height=5.2cm]{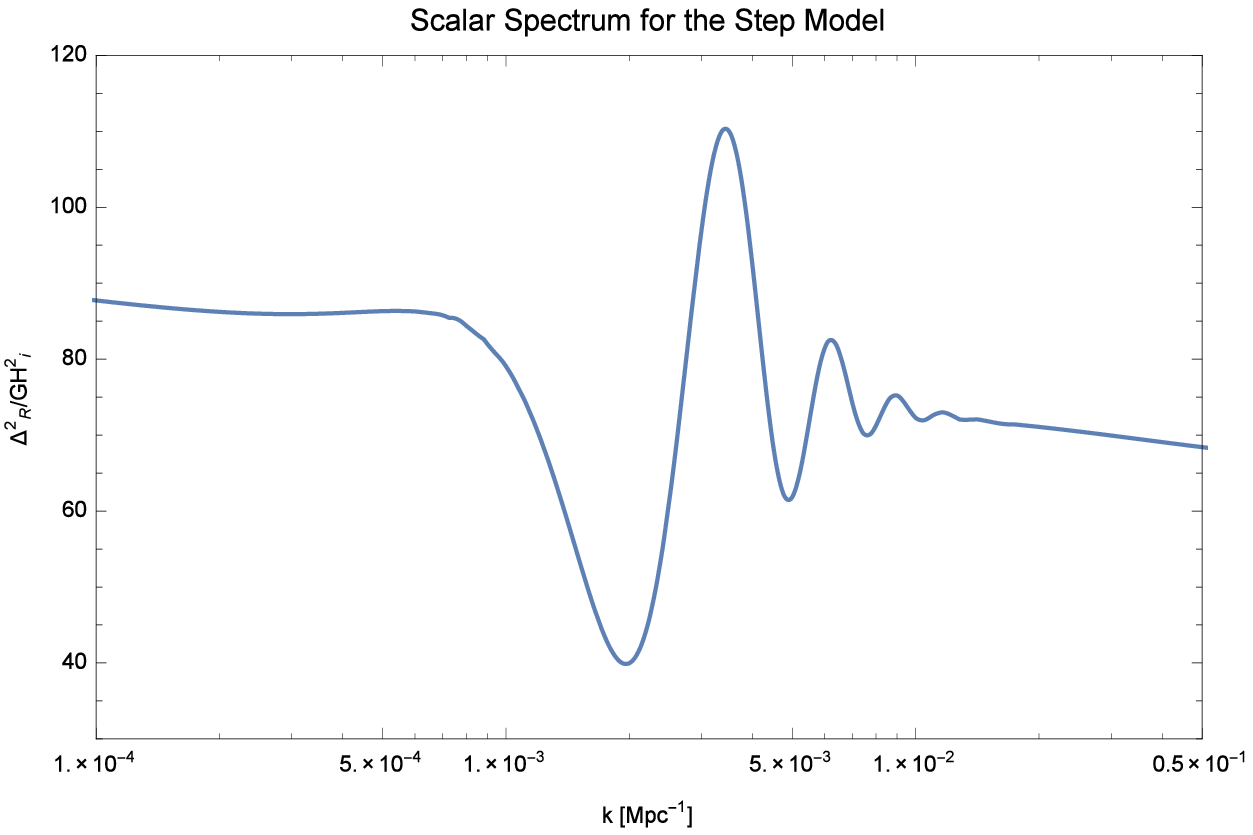}
\caption{The left hand graph represents $\Delta^2_{\mathcal{R}}(k)/G H^2_i$ 
versus $N_k$, the number of e-foldings from first horizon crossing until 
the end of inflation. The yellow curve represents the $m^2 \phi^2$ 
model while the blue curve shows the step model of section 3.1.
The right hand graph also gives $\Delta^2_{\mathcal{R}}(k)/G H^2_i$, but 
plotted as a function of the wave number $k$.}
\label{Dvkhu}
\end{figure}

\begin{figure}[ht]
\includegraphics[width=6.0cm,height=4.8cm]{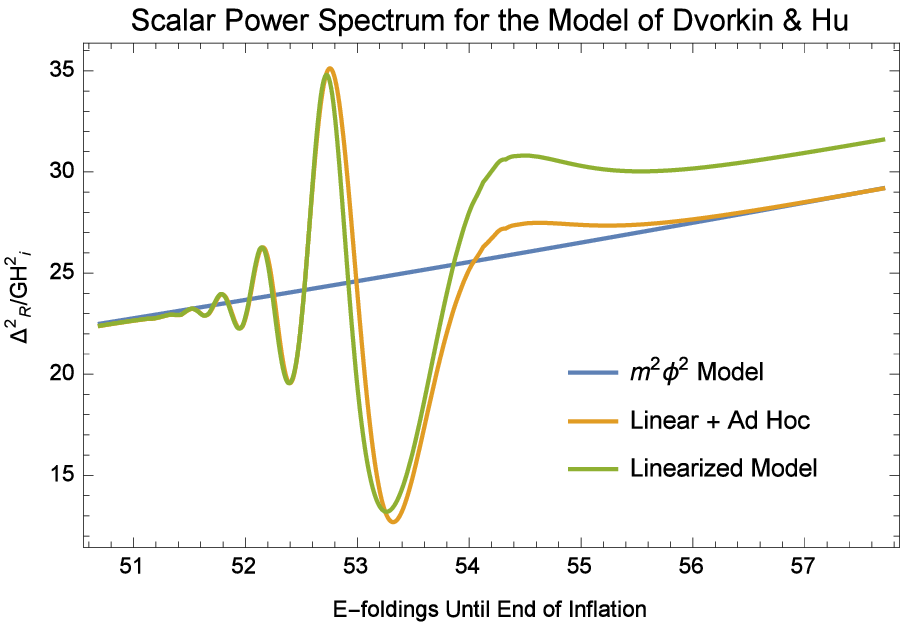}
\hspace{1cm}
\includegraphics[width=6.0cm,height=4.8cm]{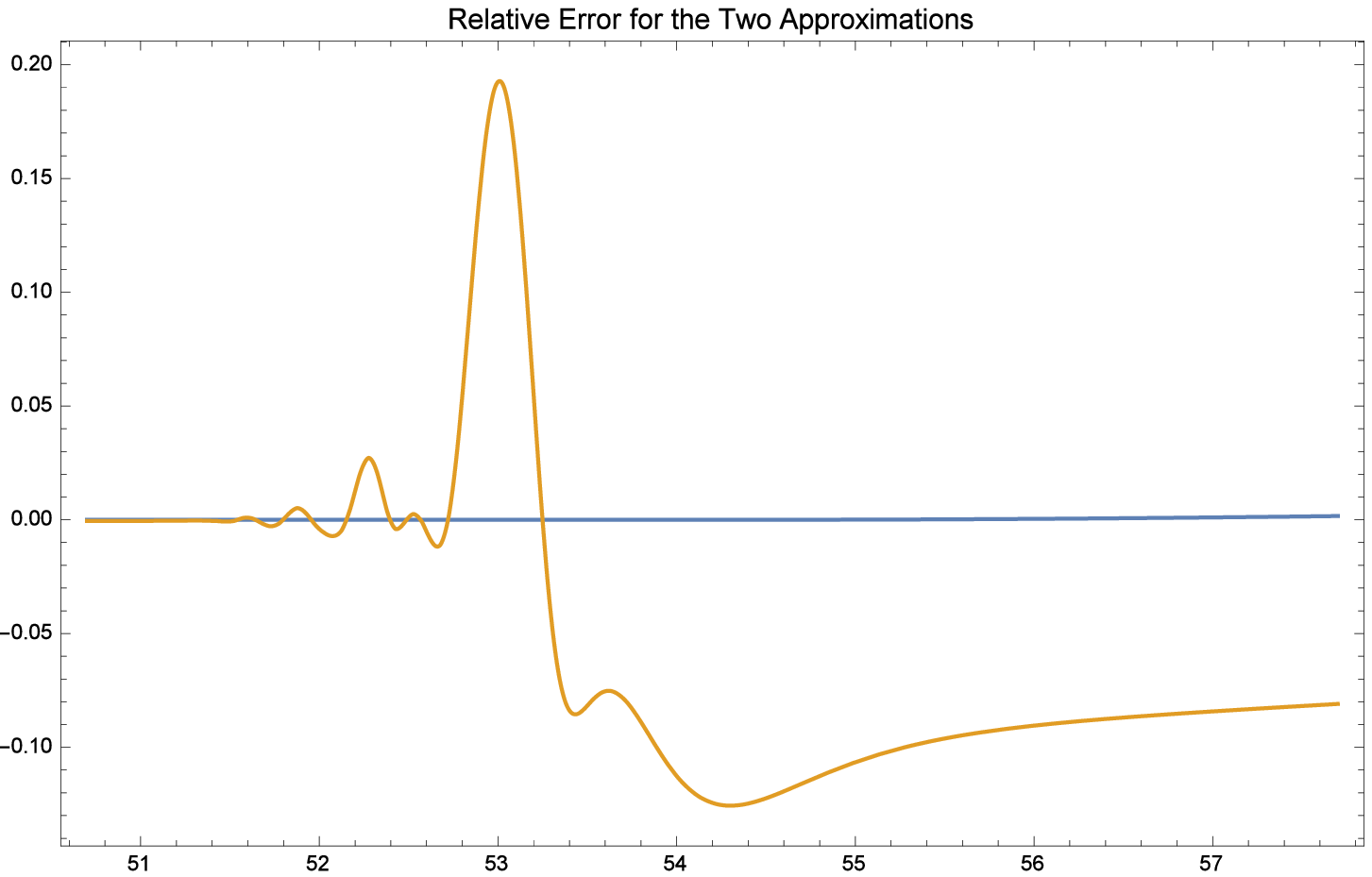}
\caption{The left hand graph shows the linearized approximation (green) and the 
modified source (yellow) for the step model of section 3.1. The right hand graph 
shows the fractional error of our linearized approximation (yellow) and the 
modified source (blue).}
\label{DvkHuApprox}
\end{figure}

\subsection{Exact Results versus Approximations}

We numerically evolved $g(n,k)$ using the exact equation (\ref{Nstep5}) 
for the step model of section 3.1. Then the $n$-independent constant 
$\mathcal{S}(k)$ was inferred by comparing the late time form with 
expression (\ref{glate}), and was used in (\ref{DRform}) to compute the 
scalar power spectrum. Figure~\ref{Dvkhu} shows the result as a function
of the number of e-foldings $N_k$ from the end of inflation that 
horizon crossing takes place. Also shown is the result for the simple
quadratic potential, without the feature. Five oscillations around the 
nominal $m^2 \varphi^2$ model are visible. They begin with a small rise 
in the power, beginning about an e-folding before the onset of the feature.
This is followed by a much larger decrease (to as low as 50\% of the $m^2
\varphi^2$ amplitude), starting in the midpoint of the feature and extending 
about half an e-folding beyond. Each subsequent fluctuation is smaller, with 
the frequency of oscillation also decreasing. The ringing persists for about 
two e-foldings after the feature has passed.

\begin{figure}[ht]
\includegraphics[width=4.0cm,height=3.0cm]{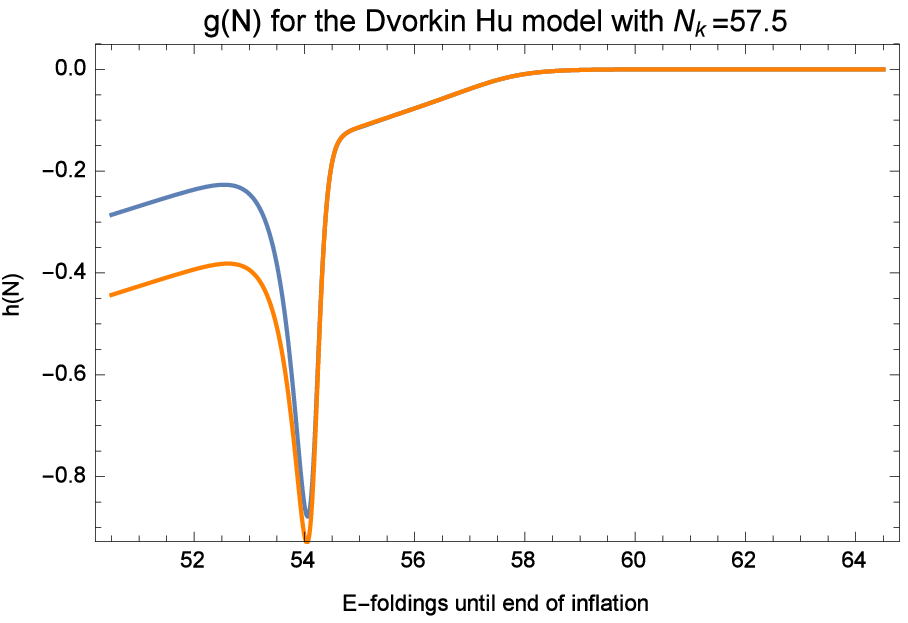}
\hspace{.6cm}
\includegraphics[width=4.0cm,height=3.0cm]{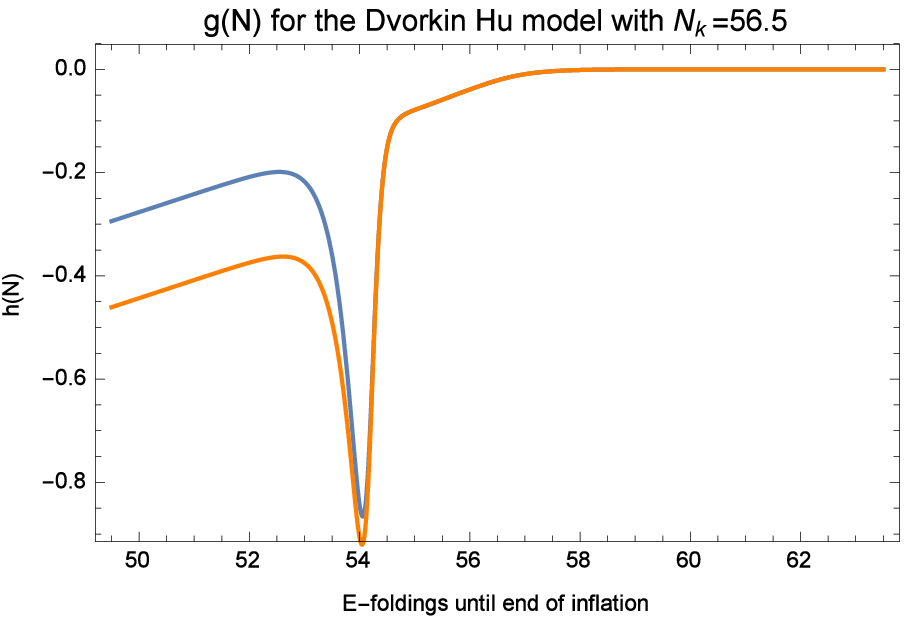}
\hspace{.6cm}
\includegraphics[width=4.0cm,height=3.0cm]{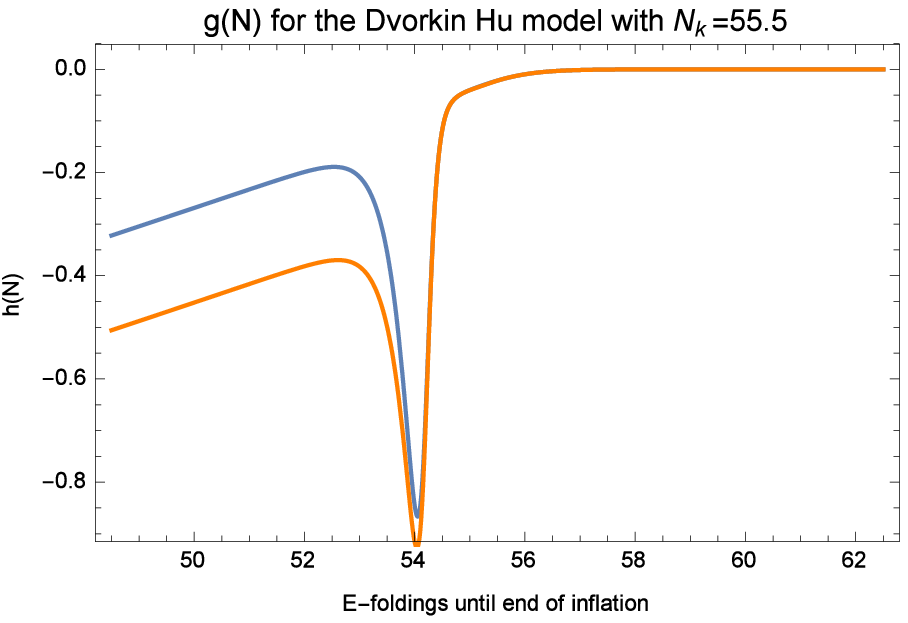}
\caption{Graphs of our linearized approximation $g(n,k) \approx g_1(n,k)$ 
(yellow) versus the full result (in blue) for the step model of section 
3.1 at very early horizon crossing times. Shown are $N_k = 57.5$ (left), 
$N_k = 56.5$ (middle) and $N_k = 55.5$ (right). After horizon crossing 
the full $g(n,k)$ assumes a downward sloping decline which the feature 
temporarily disturbs and also introduces a further, and incorrect, 
downward offset.}
\label{hveryearly}
\end{figure}

Understanding the ringing merits a sub-section of its own, but let us 
first consider the accuracy of the linearized approximation $g(n,k) 
\approx g_1(n,k)$ defined in expression (\ref{g1}). 
Figure~\ref{DvkHuApprox} displays the linearized approximation, as 
well as its fractional error. All of the oscillations are present,
at about the right places, and the 20\% maximum error is about what
Dvorkin and Hu found using a second order generalized slow roll 
expansion based on the mode functions \cite{Dvorkin:2009ne}. Also shown
in Figure~\ref{DvkHuApprox} is the effect of including a simplified
form of $g_2(n,k)$ in which only the asymptotic form (\ref{g'late}) 
for $g'(n,k)$ is used in expression (\ref{g2}). This might be termed
the ``1.5 order approximation'', and it is so good that one's eye 
cannot distinguish it from the exact curve. 

One disturbing thing of the linearized approximation is the systematic
enhancement of the power spectrum for modes which experience horizon 
crossing before the feature. This can be clearly seen in 
Figure~\ref{DvkHuApprox}. The fact that this problem disappears from 
the 1.5 order approximation signals its origin from the feature causing 
$g'(n,k)$ to become significant. Figure~\ref{hveryearly} shows what is 
going on with the full $g(n,k)$ for three modes which experience horizon 
crossing before the onset of the feature. After horizon crossing, and
before the feature, these modes have settled into the asymptotic form
predicted by expression (\ref{g'late}), with the slight downward slope
caused by the continuing growth of $\epsilon$ in the $m^2 \varphi^2$ 
model. Then the feature induces a rapid fall and rise of $g(n,k)$,
which makes the $\frac14 {g'}^2$ term on the right hand side of
(\ref{Nstep5}) impart a positive impulse to $g(n,k)$. Neglecting this
term --- which the linearized approximation does --- results in $g_1(n,k)$
being more negative than $g(n,k)$, and hence in the power spectrum of
the linearized model having a greater amplitude than it should. 

\subsection{Understanding the Ringing}

Figure~\ref{Dvkhu} of the scalar power spectrum for the step model
displays four striking properties we seek to understand:
\begin{enumerate}
\item{As the time of horizon crossing is made later, the power is 
alternatively enhanced or suppressed, with respect to the $m^2
\varphi^2$ model on which the feature was imposed, with 
progressively decreasing amplitude;}
\item{As the time of horizon crossing is made later, the period of
oscillation decreases, significantly for early horizon crossings, 
but only slightly for late horizon crossings;}
\item{Modes which experience horizon crossing near the end of the 
feature and just afterwards show a large suppression of power; and}
\item{Modes which experience horizon crossing before the onset of
the feature show a slight enhancement.}
\end{enumerate}
Of course the value of the power spectrum for a given wave number 
$k$ depends on the asymptotic form of $g(n,k)$, through relations
(\ref{glate}) and (\ref{DRform}). However, what asymptotic form is
reached depends on the previous evolution, and we can understand
that evolution by thinking about $g(n,k)$ as a damped, driven 
oscillator whose restoring force is huge before horizon crossing
and then rapidly drops away while friction persists.

\begin{figure}[ht]
\includegraphics[width=4.0cm,height=3.0cm]{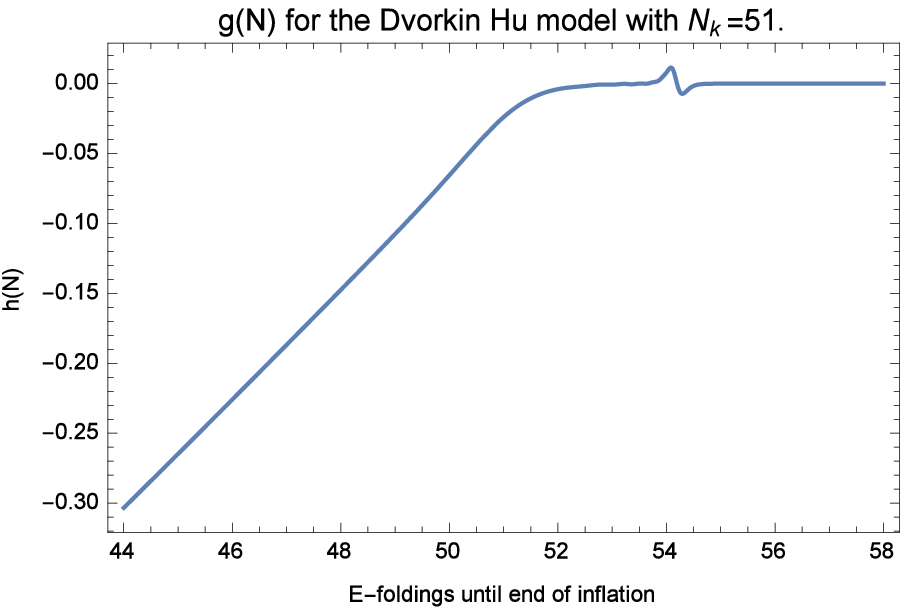}
\hspace{.5cm}
\includegraphics[width=4.0cm,height=3.0cm]{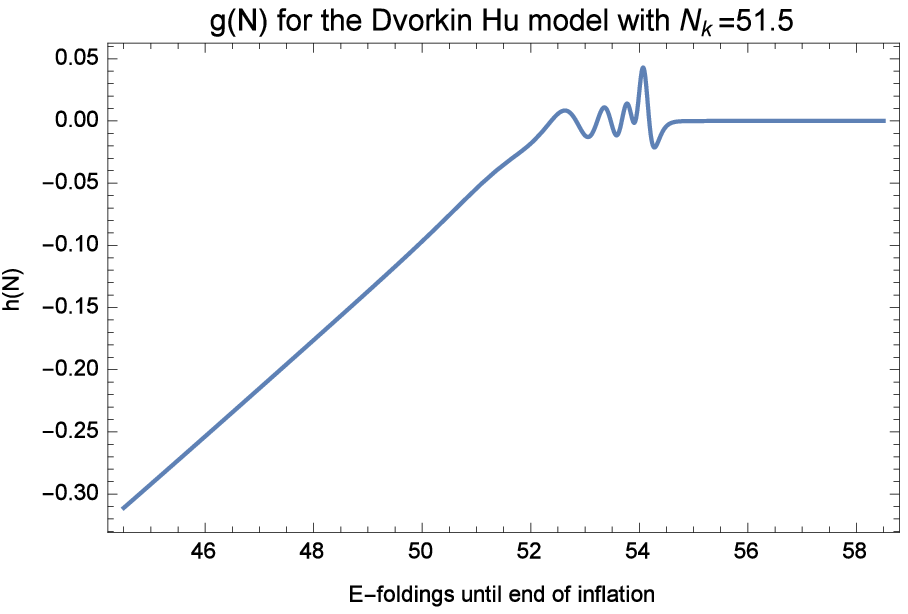}
\hspace{.5cm}
\includegraphics[width=4.0cm,height=3.0cm]{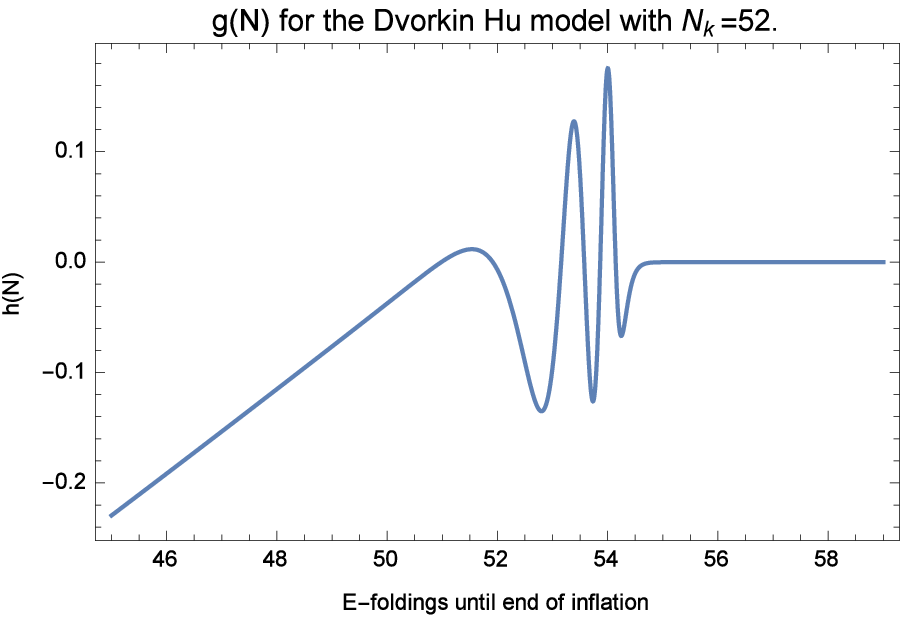}
\caption{Graphs of $g(n,k)$ for the step model of section 3.1 for 
horizon crossing times after the end of the feature. Shown are the 
evolutions (early to the right and late to the left) for horizon 
crossings at $N_k = 51.0$ (left), $N_k = 51.5$ (middle) and $N_k 
= 52.0$ (right).}
\label{hafter}
\end{figure}

The driving force consists principally of $\Delta S(n)$, which acts 
between $53.7 \lesssim N \lesssim 54.7$. Much depends on when
horizon crossing occurs with respect to that interval. 
Figure~\ref{hafter} displays $g(n,k)$ for three values of $N_k$ 
which occur after the feature. In this case the restoring force is
huge during the lifetime of the feature, so $g(n,k)$ tracks the 
driving force according to relation (\ref{gearly}). For the latest
horizon crossing ($N_k = 51.0$) there is only this single oscillation.
However, when horizon comes at $N_k = 51.5$, the restoring force has
declined and we begin to see subsequent oscillations which result 
from the oscillator responding to the net negative impulse imparted 
by $\Delta S_2(n) = -[\epsilon'/\epsilon]^2$. These oscillations
stop slightly before horizon crossing because the restoring force 
is no longer present. By $N_k = 52.0$ the restoring force is 
correspondingly smaller during the lifetime of the feature, which 
makes the amplitude of fluctuation larger. However, there is less 
time before the restoring force dies away, so there are fewer
fluctuations. One can also note that the period of oscillation
increases.

\begin{figure}
\includegraphics[width=4.0cm,height=3.0cm]{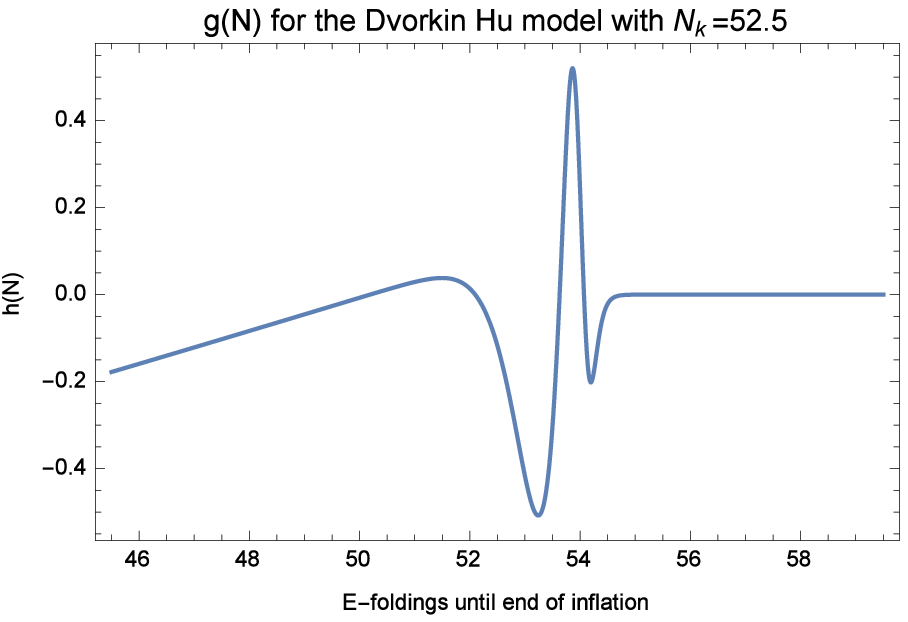}
\hspace{.5cm}
\includegraphics[width=4.0cm,height=3.0cm]{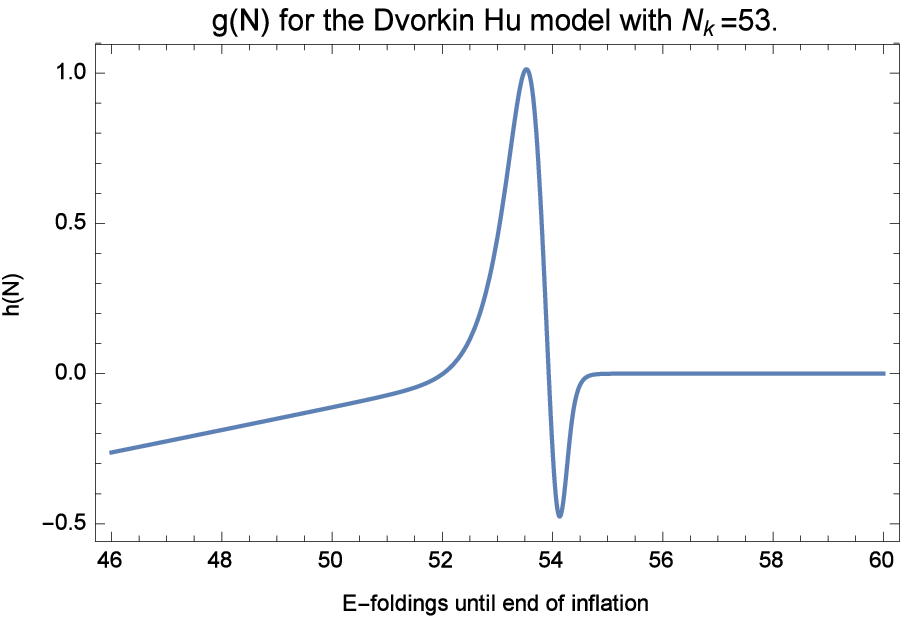}
\hspace{.5cm}
\includegraphics[width=4.0cm,height=3.0cm]{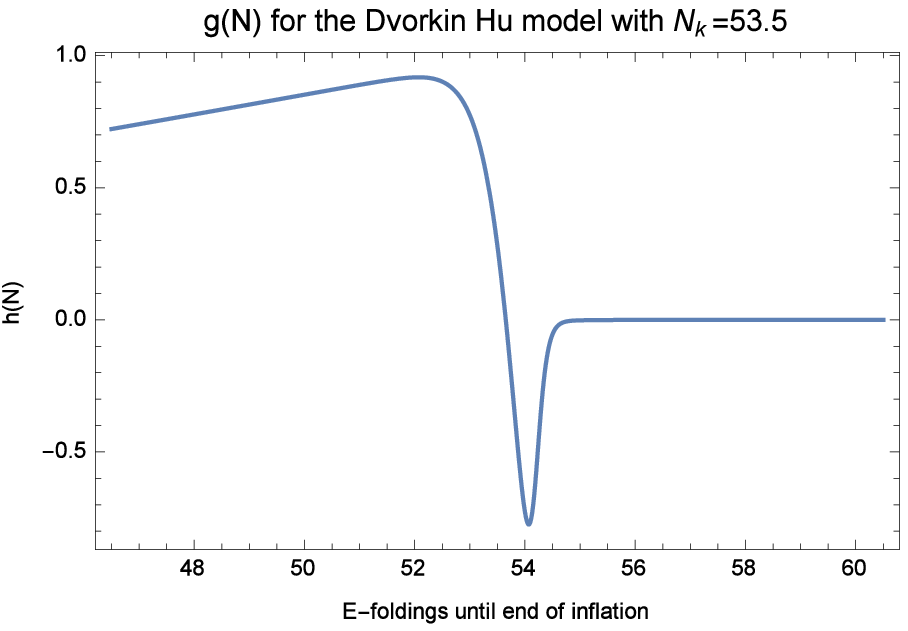}
\caption{Graphs of $g(n,k)$ for the step model of section 3.1 for 
horizon crossing times shortly after the end of the feature (at 
$N \approx 53.7$. Shown are the evolutions (early to the right 
and late to the left) for horizon crossings at $N_k = 52.5$ (left), 
$N_k = 53.0$ (middle) and $N_k = 53.5$ (right).}
\label{hduring}
\end{figure}

All of these trends are continued for the sequence of evolutions
shown in Figure~\ref{hduring} for modes which experience horizon
crossing shortly after the end of the feature. For these modes 
the restoring force is smaller during the lifetime of the feature,
which makes for fluctuations of larger amplitude. However, the
restoring force also disappears sooner, which makes for fewer
oscillations. The case of $N_k = 52.5$ shows two troughs and one
peak, while $N_k = 53.0$ has one trough and one peak, and $N_k =
53.5$ has only a single trough.

A little reflection reveals that we have the explanation for 
the first property of Figure~\ref{Dvkhu}. The value of the power 
spectrum for any $k$ depends on the asymptotic form (\ref{glate}) 
reached by $g(n,k)$. At fixed $k$, this asymptotic form is determined
by where $g(n,k)$ is in the sequence of oscillations caused by
the feature, when the restoring force turns off. For some values
of $k$ that point comes when $g(n,k)$ is in a trough, for other
values of $k$ the restoring force turns off when $g(n,k)$ is at
a peak. That is why there are fluctuations. The amplitude of
fluctuation becomes smaller for later horizon crossing because
the restoring force at the time of the feature is correspondingly
larger.

The second property, concerning the period of oscillation, has a 
more complicated explanation associated with the nonlinear terms in 
equation (\ref{Nstep5}) for $g(n,k)$. The actual restoring force is
$\omega^2 [e^g - 1]$, so positive values of $g(n,k)$ increase it, 
with respect to the linear approximation, while negative values 
of $g(n,k)$ decrease it. The system tends to spend a little less
time on upward fluctuations, and a little more on downwards ones,
with the net effect a lengthening of the period for one full
oscillation. One can see from Figure~\ref{hduring} that the 
amplitude of $g(n,k)$ becomes large enough to make the
oscillator significantly anharmonic. The later horizon crossing
times of Figure~\ref{hafter} never reach such large amplitudes,
which is why the period of fluctuations on Figure~\ref{Dvkhu}
decreases as $N_k$ becomes smaller.

Property 3, the large enhancement for modes which experience
horizon crossing near the end of the feature, is shown by the 
final graph on Figure~\ref{hduring}. For $N_k = 53.5$ the
restoring force is of order one just a little after the feature
has ended. The restoring force turns off just as the system is
rebounding, whereupon it continues to drift upwards until stopped
by friction. Positive $g(n,k)$ means $\Delta \mathcal{N}(t,k) =
\exp[-\frac12 g(n,k)]$ is decreased, so the power should be 
reduced, which is what one sees on Figure~\ref{Dvkhu}.

\begin{figure}[ht]
\includegraphics[width=4.0cm,height=3.0cm]{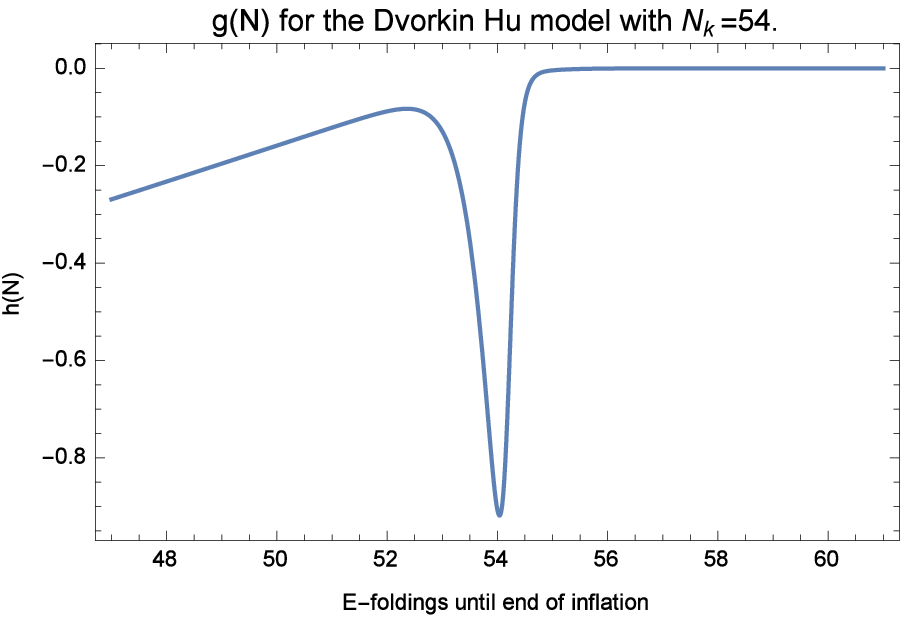}
\hspace{.5cm}
\includegraphics[width=4.0cm,height=3.0cm]{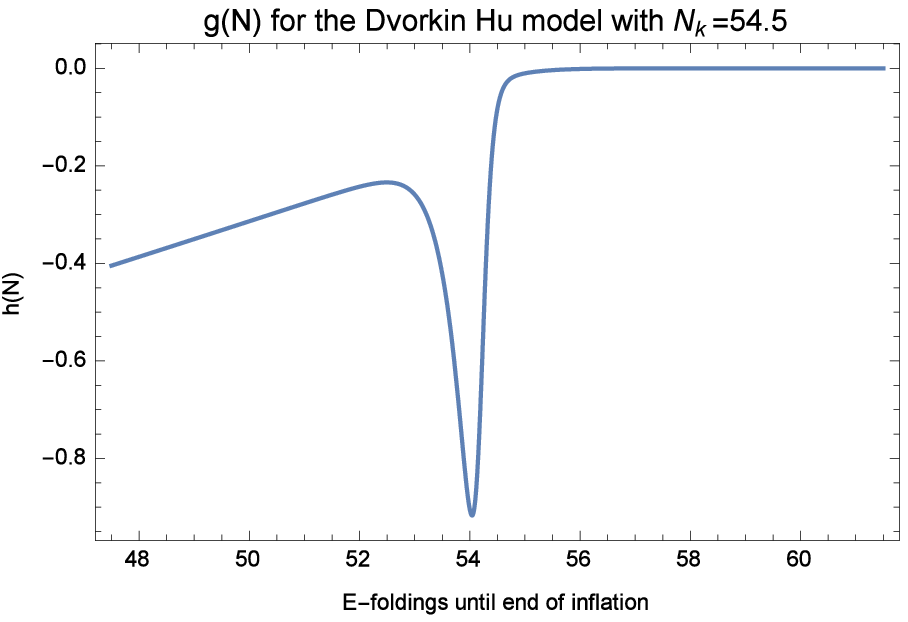}
\hspace{.5cm}
\includegraphics[width=4.0cm,height=3.0cm]{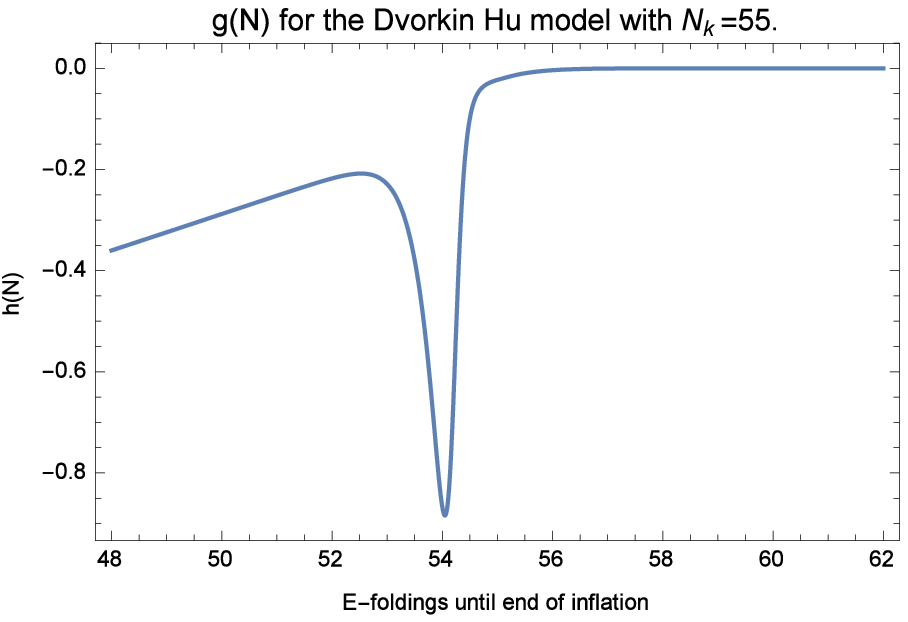}
\caption{Graphs of $g(n,k)$ for the step model of section 3.1 for 
horizon crossing times during and slightly before the onset of the 
feature. Shown are the evolutions (early to the right and late to 
the left) for horizon crossings at $N_k = 54.0$ (left), $N_k = 54.5$ 
(middle) and $N_k = 55.0$ (right).}
\label{hbefore}
\end{figure}

Figure~\ref{hbefore} shows three modes whose horizon crossing times
occur during and slightly before the feature. For these modes the 
restoring force is nearly absent during the lifetime of the feature 
so what one sees is the effect of the driving force on a massless 
particle with friction. The driving force $\Delta S(n)$ (\ref{Nsource})
is depicted on the right hand side of Figure~\ref{modelB}. Without 
friction it would first push $g(n,k)$ negative and then pull it back 
up. With friction, and the small negative impulse imparted by $\Delta 
S_2(n)$ (\ref{DS12}), one can see that $g(n,k)$ will not quite reach 
equilibrium, and this is reflected in Figure~\ref{hbefore}. The result 
is to make $g(n,k)$ more negative than it would have been without the 
feature, which increases $\Delta \mathcal{N}(t,k) = \exp[-\frac12 
g(n,k)]$. The net effect on the power spectrum is decreased because 
these modes come during the lifetime of the feature. One can see from 
Figure~\ref{modelA} that the factor of $1/\epsilon(t_k)$ in 
(\ref{DRform}) is decreased. That is why the peak on the right of 
Figure~\ref{Dvkhu} is shallow.

\subsection{The Impulse Model}

\begin{figure}[ht]
\includegraphics[width=6.0cm,height=4.8cm]{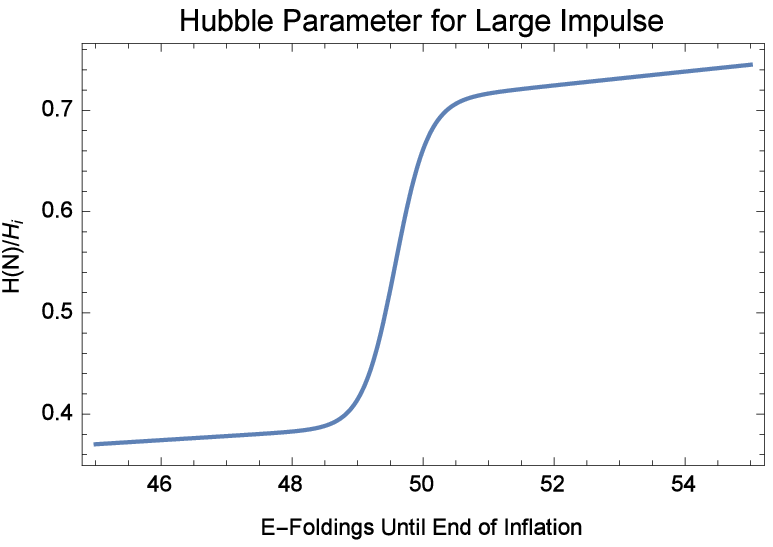}
\hspace{1cm}
\includegraphics[width=6.0cm,height=4.8cm]{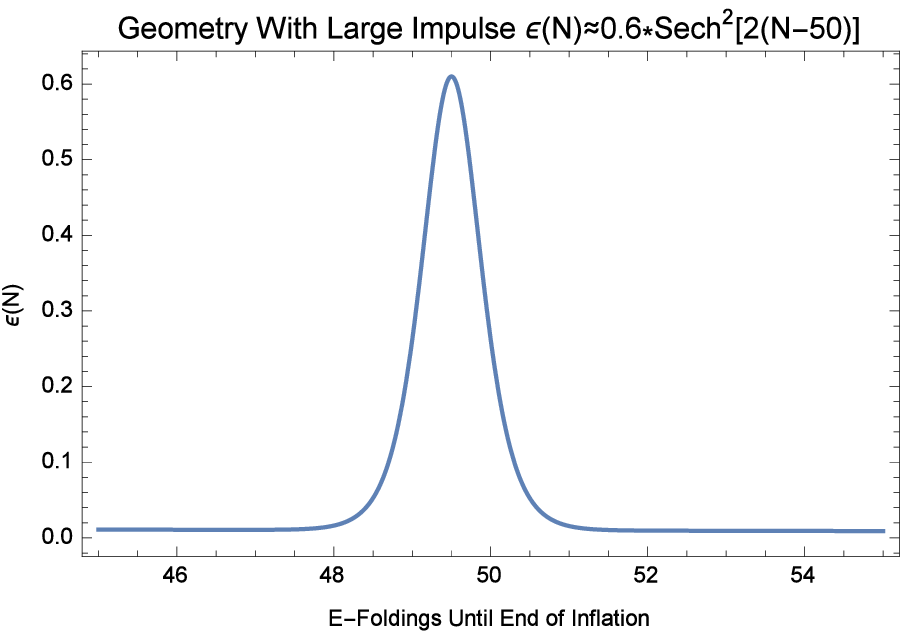}
\caption{Graphs of the Hubble parameter and the first slow roll 
parameter as functions of the number of e-foldings $N \equiv n_{\rm end}
- n$ until the end of inflation for the impulse model of section 3.4. Note 
that the passage of the feature causes a large reduction in the Hubble
parameter.}
\label{altmodelgeom}
\end{figure}

Recall that the extra scalar source (\ref{Nsource}) can be decomposed 
into two parts (\ref{DS12}): a total derivative $\Delta S_1(n)$ and
a negative-definite part $\Delta S_2(n)$ which contributes the net
impulse. is negative-definite. The step model typifies features for 
which $\vert \Delta S_1(n) \vert \gg \vert \Delta S_2(n) \vert$, and 
the response we saw in sections 3.2 and 3.3 is generic. It is difficult 
to make $\Delta S_2(n)$ dominate $\Delta S_1(n)$ for very long, but 
there is another class of features for which the two terms are 
comparable, and this class shows a very different response.

\begin{figure}[ht]
\includegraphics[width=10.0cm,height=8.0cm]{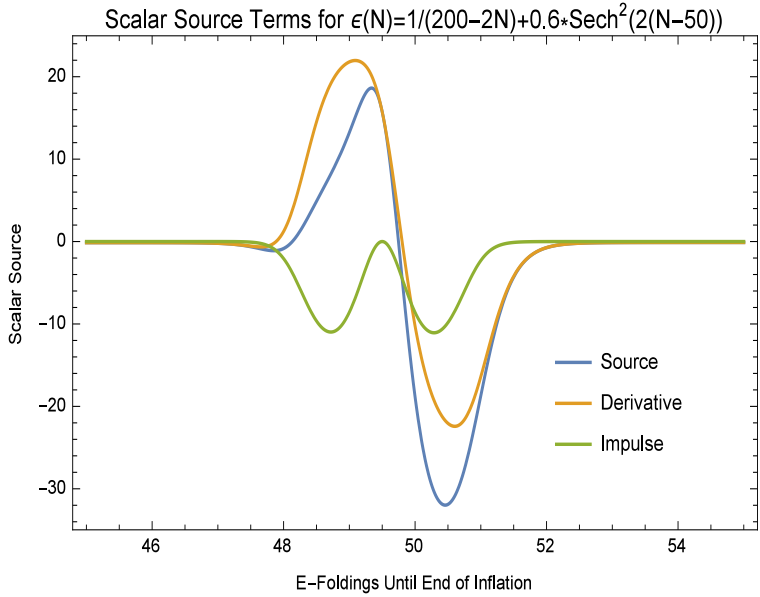}
\caption{Graphs of the scalar source for the model 
(\ref{altmodeleps}-\ref{altmodelH}) of section 3.4. Shown are 
$\Delta S(n)$ (blue), $\Delta S_1(n)$ (orange) and $\Delta S_2(n)$ 
(green).}
 \label{Impulsesource}
\end{figure}

One can make $\Delta S_1(n)$ and $\Delta S_2(n)$ comparable by
causing the peak value of $\epsilon$ at the feature to be much larger 
than the background on which it is imposed. As an example, consider
the model defined by,
\begin{equation}
\epsilon(n) = \frac1{200 \!-\! 2n} + \frac35 \, {\rm sech}^2\Bigl[2 
(n \!-\! 50)\Bigr] \; . \label{altmodeleps}
\end{equation}
The associated Hubble parameter is,
\begin{equation}
H(n) = H_i \sqrt{1 \!-\! \frac{n}{100}} \, \exp\Biggl[-\frac3{10} \!-\! 
\frac3{10} \tanh\Bigl[2 (n\!-\!50)\Bigr]\Biggr] \; . \label{altmodelH}
\end{equation}
Figure~\ref{altmodelgeom} displays (\ref{altmodeleps}-\ref{altmodelH}).
Note that the feature effectively persists from $N \equiv n_{\rm end} - n
\simeq 51.5$ to $N \simeq 48.0$, so it is much broader than the step
model. Note also that the passage of the feature reduces $H^2$ by about
a factor of two, whereas the step model of section 3.1 shows hardly any 
change. Figure~{\ref{Impulsesource} displays $\Delta S(n)$ and its two 
parts. Note that the magnitude of $\Delta S_2$ is only a factor of two 
smaller than that of $\Delta S_1$, whereas the magnitudes differ by a 
factor of ten for the step model.

\begin{figure}[ht]
\includegraphics[width=4cm,height=3cm]{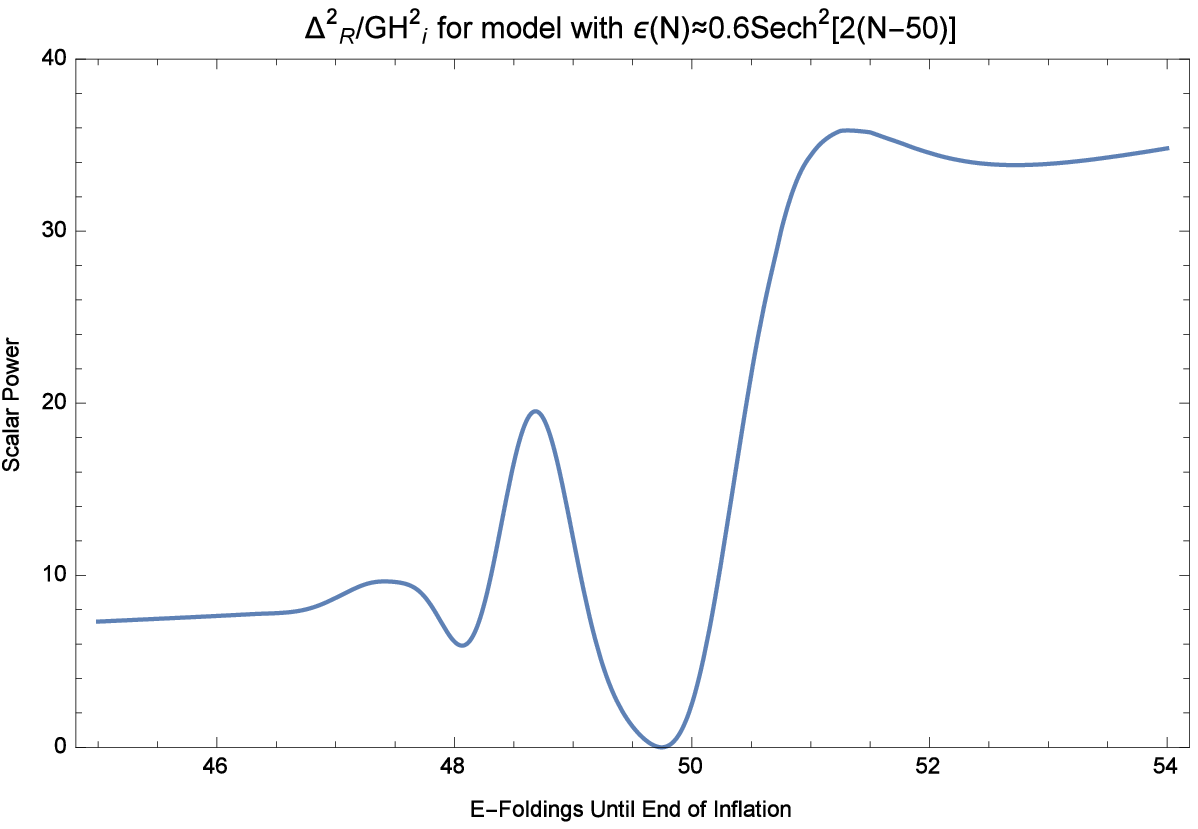}
\hspace{.6cm}
\includegraphics[width=4cm,height=3cm]{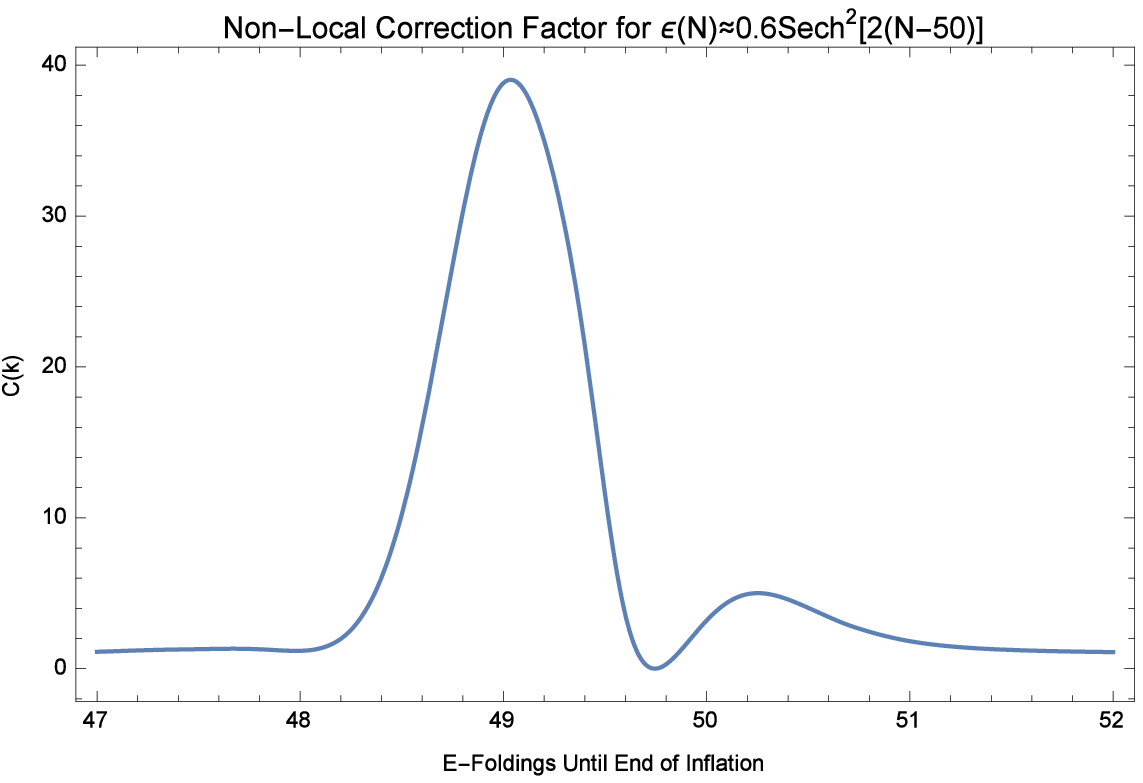}
\hspace{.6cm}
\includegraphics[width=4cm,height=3cm]{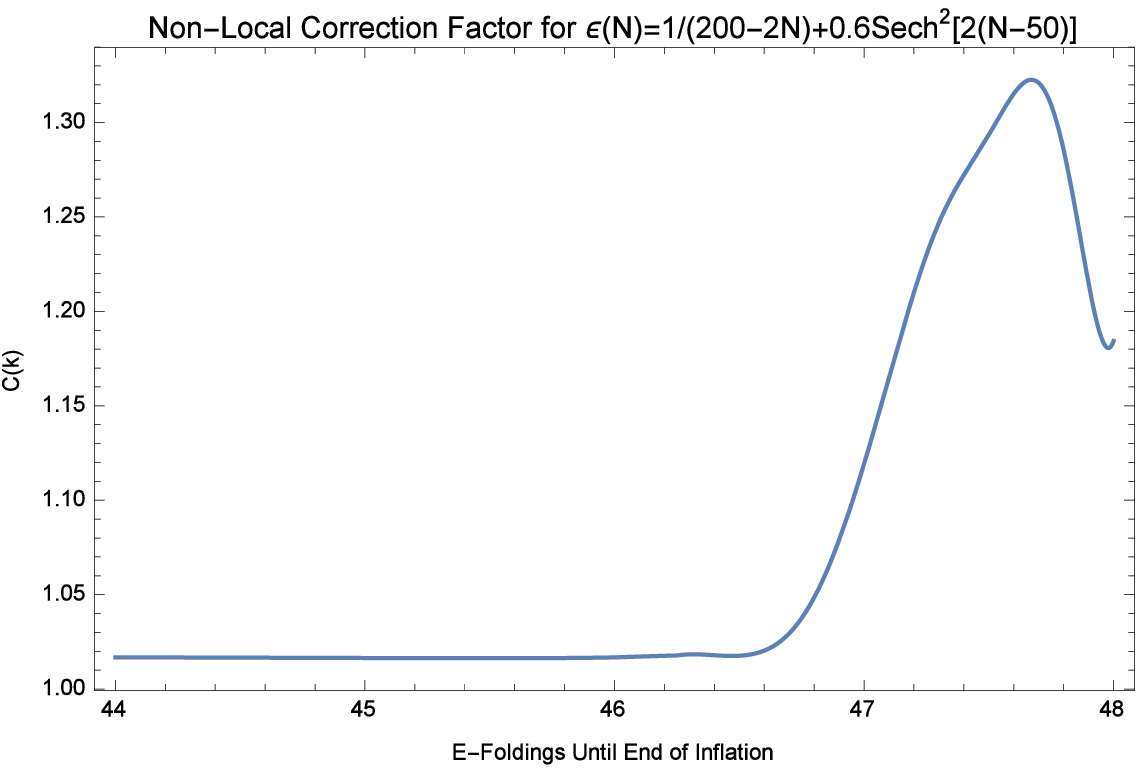}
\caption{The left hand graph shows $\Delta^2_{\mathcal{R}}(k)/G H_i^2$ 
for the impulse model of section 3.4. The middle and right hand graphs 
show the nonlocal factor $\mathcal{S}(k)$ for this model. All graphs are 
displayed as function of $N_k$, the number of e-foldings between horizon 
crossing and the end of inflation.}
\label{Impulsepower}
\end{figure}

Figure~\ref{Impulsepower} shows the scalar power spectrum for the
impulse model. Also displayed is the nonlocal factor $\mathcal{S}(k)$ 
from our representation (\ref{DRform}). Recall that $\mathcal{S}(k)$ 
is the part of the power spectrum which is not attributable to local 
changes of the geometry at the time of horizon crossing. The feature 
is so big in this model that the largest fluctuation of $\Delta^2_{
\mathcal{R}}(k)$ derives from changes in the leading slow roll 
approximation, $G H^2(t_k)/\pi \epsilon(t_k)$. That is what suppresses 
the power so much in the range $49.0 \lesssim N_k \lesssim 50.5$.
However, there are also quite large effects from $\mathcal{S}(k)$,
including the big bump in the region $48.5 \lesssim N_k \lesssim 49.5$ 
and the smaller bump in the range $50.0 \lesssim N_k \lesssim 50.7$. 
The feature of this model is so broad that much of its impulse is 
dissipated during the feature, and subsequent ringing is less than
for the step model.

\begin{figure}[ht]
\includegraphics[width=6.0cm,height=4.8cm]{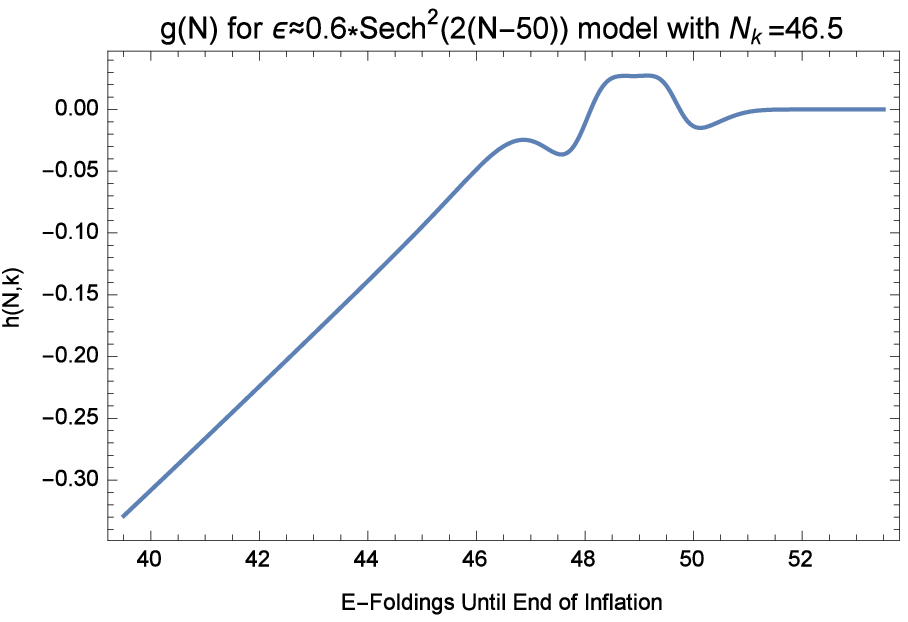}
\hspace{1cm}
\includegraphics[width=6.0cm,height=4.8cm]{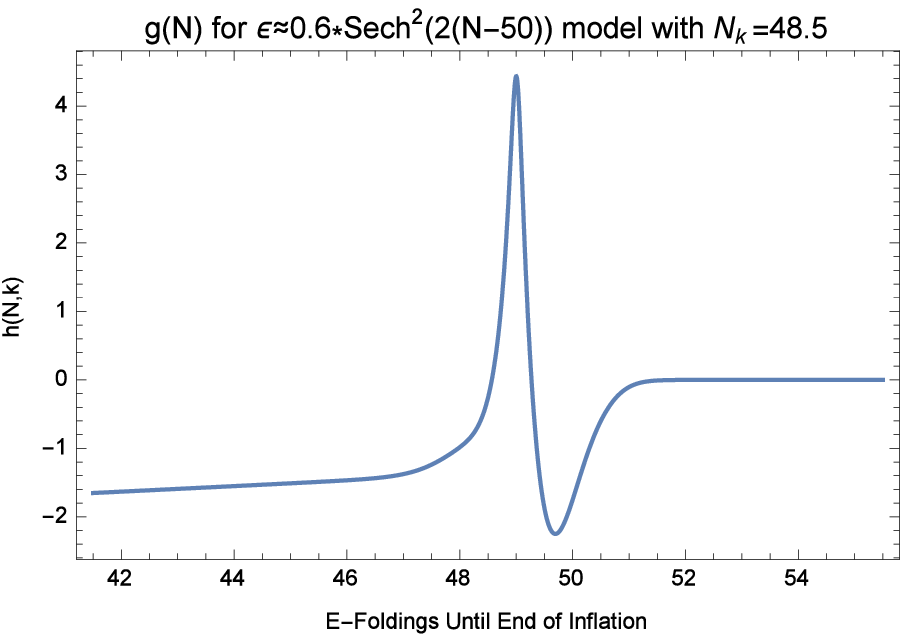}
\caption{Graphs of $g(n,k)$ for the impulse model of section 3.4 for
two modes which experience horizon crossing after (on the right) and
near the end of (on the left) the feature. Both curves are displayed 
as functions of $N \equiv n_{\rm end} - n$, the number of e-foldings 
until the end of inflation.}
\label{Impulseafter}
\end{figure}

Figure~\ref{Impulseafter} shows the evolution of $g(n,k)$ for two modes
which experience horizon crossing after, and near the end of, the feature.
In the first case ($N_k = 46.5$) the restoring force is still significant
during the feature, which suppresses the response. It is worthwhile 
comparing this evolution with that depicted on the far right of 
Figure~\ref{hafter} for a mode which experiences horizon crossing about
the same number of e-foldings after the feature of the step model. 
Although $g(n,k)$ is quite different during the feature --- because the
sources are different --- the subsequent evolution is quite similar. For
the second of the Figure~\ref{Impulseafter} graphs ($N_k = 48.5$) the 
initial, downward push from the source is suppressed by the still 
significant restoring force, but the final, upward push is unsuppressed 
as the restoring force dissipates. The large values of $g(n,k)$ mean 
that nonlinear effects are significant. 

\begin{figure}[ht]
\includegraphics[width=6.0cm,height=4.8cm]{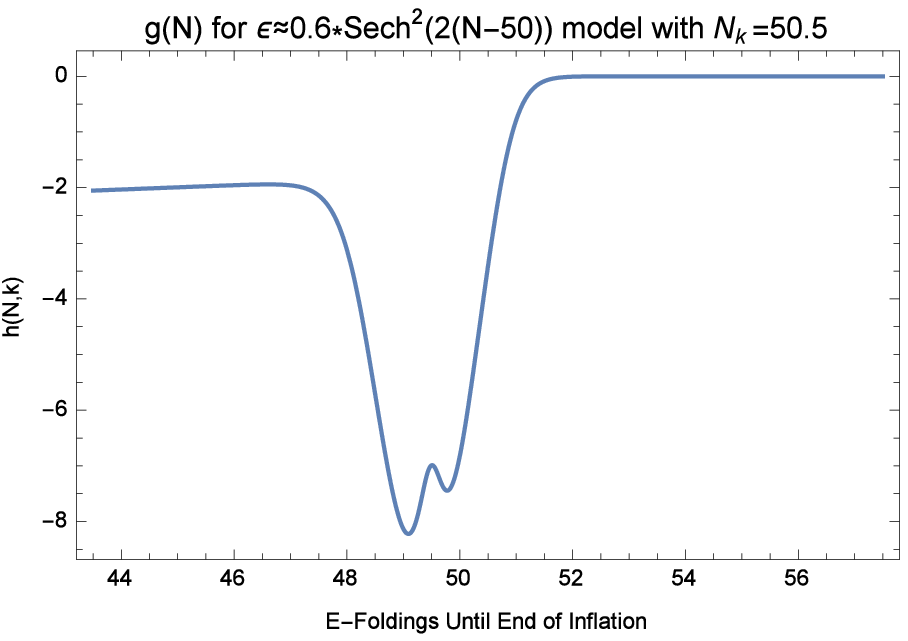}
\hspace{1cm}
\includegraphics[width=6.0cm,height=4.8cm]{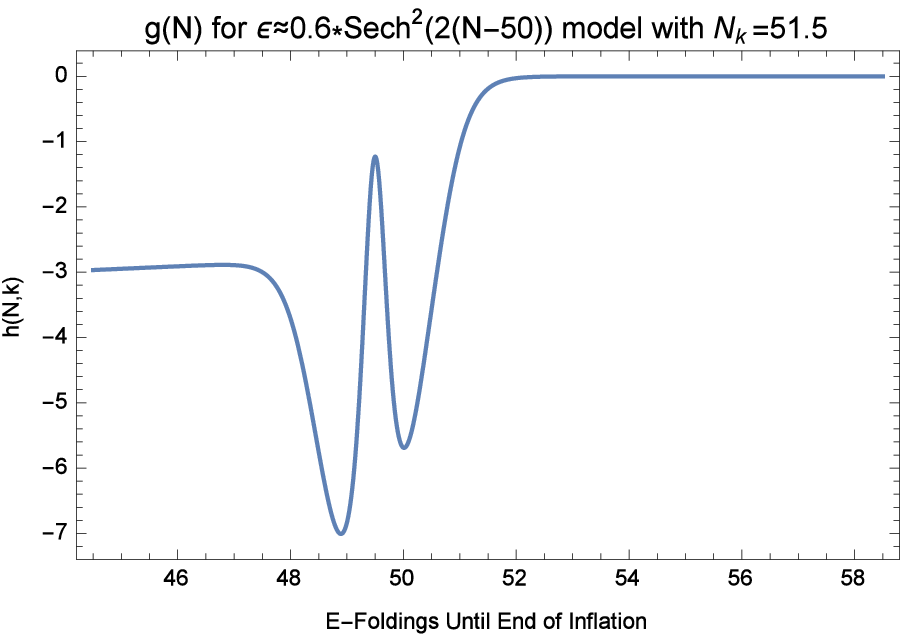}
\caption{Graphs of $g(n,k)$ for the impulse model of section 3.4 for
two modes which experience horizon crossing during the passage of the
feature. Both curves are displayed as functions of $N \equiv n_{\rm end}
-n$, the number of e-foldings until the end of inflation.}
\label{Impulsebefore}
\end{figure}

Figure~\ref{Impulsebefore} shows $g(n,k)$ for two modes which experience
horizon crossing during the first part of the feature. In both cases 
$g(n,k)$ is pushed in the negative direction due to the net negative 
impulse which is imparted by $\Delta S_2(n)$. Because the feature of 
the impulse model is so much larger than that of the step model, both
of these modes are pushed into the nonlinear regime. In both cases, one
can note rebounds at $N = 49.5$, corresponding to where $\Delta S_2(n)$
vanishes. For the earlier horizon crossing ($N_k = 51.5$) this rebound
is quite prominent, presumably because the restoring force is small. 

\section{Epilogue}

Our goal is to understand how the scalar and tensor power spectra
depend {\it functionally} on the expansion history of inflation.
That is, we wish to express them as functionals of $a(t)$ for a 
general inflationary expansion history. A previous study 
\cite{Brooker:2015iya} developed such a representation for the 
tensor power spectrum based on first deriving an evolution equation
(\ref{Meqn}) for the norm-squared of the tensor mode function
\cite{Romania:2012tb}, then factoring out the instantaneously 
constant $\epsilon$ solution (\ref{Mstep1}), and finally recognizing
that the exact equation (\ref{Mstep5}) for the residual factor has
a series solution based on an explicit Green's function (\ref{Gfunct}).
This expansion is so good that only the first term (\ref{h1}) is
needed for normal models \cite{Brooker:2015iya}.

The goal of this paper was to develop a functional representation for
the scalar power spectrum. Two competing strategies were described in
section 2: either exploit the exact functional relation (\ref{MtoN})
between the tensor and scalar power spectra, or else parallel the 
tensor analysis by factoring out the instantaneously constant $\epsilon$
solution. Of course the transformation (\ref{MtoN}) is exact, and would
be effective if used on the exact tensor power spectrum. However, our
numerical studies of section 3 show that it does not give accurate
results when used on our linearized approximation. The reason is that
the presence of even small features in the original expansion history 
induces wild behavior in the transformed geometry required to employ
relation (\ref{MtoN}). From Figure~\ref{modelC} one can see that even 
models which have been fitted to real data \cite{Mortonson:2009qv} 
cause the transformed geometry to oscillate between normal acceleration
($0 \leq \widetilde{\epsilon} < 1$ to deceleration ($1 < \widetilde{\epsilon}$)
and even super-acceleration ($\widetilde{\epsilon} < 0$). Although our 
analytic approximation for the tensor power spectrum is very good 
\cite{Brooker:2015iya}, it breaks down for such violent fluctuations.

The more effective strategy is factoring out the instantaneously 
constant $\epsilon$ solution as described in section 2.2. One surprise 
is that the final evolution equation (\ref{Nstep5}) differs from its 
tensor cousin (\ref{Mstep5}) only by the addition of a few terms 
(\ref{Nsource}). In particular, the Green's function (\ref{Gfunct})
is identical, and the series solution (\ref{g1}-\ref{g2}) differs 
only by the inclusion of this extra source in the first term. This 
allowed us to largely read off the asymptotic behaviors (\ref{gearly}) 
and (\ref{glate}) from our previous tensor study \cite{Brooker:2015iya}.

$\Delta^2_{\mathcal{R}}(k)$ takes the form (\ref{DRform}) of a leading
slow roll approximation, times a function (\ref{Cdef}) of the slow
roll parameter at horizon crossing, times a nonlocal functional
$\mathcal{S}(k)$ which depends upon times before and after horizon
crossing. One computes this last factor by solving for the residual 
$g(n,k)$ --- either numerically from (\ref{Nstep5}), or else by using
our series expansion (\ref{g1}-\ref{g2}) --- then comparing with the
late time asymptotic form (\ref{glate}).

It is useful to think about the scalar residual $g(n,k)$ as a damped,
driven oscillator in the same way we previously developed for the tensor 
residual $h(n,k)$ \cite{Brooker:2015iya}. Comparison of the tensor
equation (\ref{Mstep5}) with its scalar cousin (\ref{Nstep5}) reveals
that both residuals experience the same restoring force and the same
friction, and both oscillators are disrupted by the same nonlinear
terms. The initial restoring force is exponentially huge, becomes of 
order one at horizon crossing, and dies away after that with 
exponential swiftness. In contrast, the friction term is always of 
order one. This means that the driving force can only become effective
a few e-foldings before horizon crossing. If it gives the system a
sufficiently large kick during that time, there will be a series of
oscillations of decreasing frequency as the restoring force dies away
and friction damps the motion. However, it is important to understand
that the power spectra do not directly reflect these oscillations, 
only their effect on the asymptotic amplitudes of the mode functions.

The only difference between the scalar equation (\ref{Nstep5}) and
the tensor one (\ref{Mstep5}) is the presence of an extra driving force
$\Delta S(n)$ for the scalar. Both the tensor driving force $S(n,k)$
and the extra scalar part $\Delta S(n)$ are proportional to the first 
and second derivatives of the first slow roll parameter $\epsilon$.
However, $\Delta S(n)$ tends to be much larger because it involves
factors of $1/\epsilon$ which $S(n,k)$ lacks. This means that the
scalar response to a feature is much larger than the tensor response.
It also comes sooner because $S(n,k)$ is negligible before horizon 
crossing, which is not necessarily true for $\Delta S(n)$. 

Two classes of models differ by the relation between the two parts 
(\ref{DS12}) of the scalar source (\ref{Nsource}). For the step model 
of section 3.1 the magnitude of $\Delta S_1(n)$ is about ten times 
that of $\Delta S_2(n)$, whereas the two parts are comparable, and
the feature is much broader, for the impulse model of section 3.4.
The resulting, very different, power spectra are shown for the step
model in Figure~\ref{Dvkhu} and for the impulse model in 
Figure~\ref{Impulsepower}.

Although the power spectra depend only on the norm-squared of the 
mode functions, other quantities depend upon the phase, such as
non-Gaussianities and the propagators for the $h_{ij}$ and $\zeta$ 
fields. We might define the full tensor and scalar mode functions
as,
\begin{equation}
u(t,k) \equiv \sqrt{M(t,k)} \!\times\! e^{-i\theta(t,k)} \qquad ,
\qquad v(t,k) \equiv \sqrt{\mathcal{N}(t,k)} \!\times\! 
e^{-i \phi(t,k)} \; .
\end{equation}
Our work so far has been to develop good analytic expressions for
$M(t,k)$ and $\mathcal{N}(t,k)$, but it is simple to use the 
Wronskians to express the phases in terms of the magnitudes,
\begin{eqnarray}
u \dot{u}^* - \dot{u} u^* = \frac{i}{a^3} \qquad & \Longrightarrow
& \qquad \dot{\theta}(t,k) = \frac1{2 a^3(t) M(t,k)} 
\; , \label{theta} \\
v \dot{v}^* - \dot{v} v^* = \frac{i}{\epsilon a^3} \qquad &
\Longrightarrow & \qquad \dot{\phi}(t,k) = \frac1{2 \epsilon(t)
a^3(t) \mathcal{N}(t,k)} \; . \label{phi}
\end{eqnarray}
Expressions (\ref{theta}-\ref{phi}) mean that no extra work is 
needed to obtain the full mode functions. They also explain the early
time expansions for $M(t,k)$ and $\mathcal{N}(t,k)$ are local (for 
example, equation (\ref{gearly})) while the phases are not (for example,
the right hand sides of equations (\ref{ueqns}-\ref{veqns})). And 
they limit the degree of nonlocality --- in addition the that which 
may be present in the magnitudes --- to just a single integral. That
is especially important for the products of mode functions at 
different times which enter both the propagator and the tree order 
non-Gaussianity,
\begin{eqnarray}
u(t,k) u^*(t',k) & = & \sqrt{M(t,k) M(t',k)} \, \exp\Bigl[ -i \!\!
\int_{t'}^{t} \!\! \frac{dt''}{2 a^3(t'') M(t'',k)}\Bigr] \; , \\
v(t,k) v^*(t',k) & = & \sqrt{\mathcal{N}(t,k) \mathcal{N}(t',k)} \, 
\exp\Bigl[ -i \!\! \int_{t'}^{t} \!\! \frac{dt''}{2 \epsilon(t'') 
a^3(t'') \mathcal{N}(t'',k)} \Bigr] \; . \qquad 
\end{eqnarray}

Finally, we return to a question about the $\zeta$ propagator 
$i\Delta(t_1,\vec{x}_1;t_2,\vec{x}_2)$ which was raised in our tensor 
study \cite{Brooker:2015iya}. One can tell from the case of constant 
$\epsilon(t)$ that this propagator must contain a factor of 
``$1/\epsilon$''. However, it has never been clear at what time, or 
times, that factor should be evaluated. In particular, might it come 
from from some time far to the past of $t_1$ and $t_2$, when 
$\epsilon(t)$ was much smaller than either $\epsilon(t_1)$ or
$\epsilon(t_2)$? For the sake of simplicity let us assume $t_1 > t_2$,
in which case the $\zeta$ propagator is,
\begin{equation}
i\Delta(t_1,\vec{x}_1;t_2,\vec{x}_2) = \int \!\! \frac{dk}{k}
\frac{\sin(k \Delta x)}{k \Delta x} \!\times\! \frac{k^3}{2 \pi^2}
\, v(t_1,k) v^*(t_2,k) \; , \label{zetaprop}
\end{equation}
where $\Delta x \equiv \Vert \vec{x}_1 - \vec{x}_2\Vert$.
For sub-horizon wave numbers $k > H(t) a(t)$ our analysis shows that
the amplitude of the full mode function $v(t,k)$ closely tracks that
of the instantaneously constant $\epsilon$ solution $v_0(t,k;\epsilon(t))$
(\ref{constepsv}). Expression (\ref{gearly}) also shows that even the
{\it corrections} to $v_0(t,k;\epsilon(t))$ are both exponentially
suppressed and {\it local} for sub-horizon modes. Memories of earlier 
times only begin to have an effect a few e-foldings before horizon 
crossing.

Super-horizon modes behave quite differently. Their amplitudes freeze 
into to the values they held near the time of first horizon crossing,
\begin{equation}
k \ll H(t_2) a(t_2) \quad \Longrightarrow \quad \frac{k^3}{2 \pi^2} 
v(t_1,k) v^*(t_2,k) \approx \frac{H^2(t_k)}{4 \pi^2 \epsilon(t_k)} 
\!\times\! C\Bigl(\epsilon(t_k)\Bigr) \!\times\! \mathcal{S}(k) \; .
\label{super}
\end{equation} 
Of course the propagator (\ref{zetaprop}) consists of a sum over all 
modes, so the super-horizon modes can indeed depend upon times far in 
the past of either $t_1$ or $t_2$, when $\epsilon(t_k)$ was smaller
and $H(t_k)$ was larger. Because $C(\epsilon)$ is a monotonically
decreasing function of $\epsilon$, smaller $\epsilon(t_k)$ makes this
factor closer to its maximum value of one \cite{Brooker:2015iya}. Our
analysis shows that the nonlocal factor $\mathcal{S}(k)$ can indeed
be significant if the model possesses a feature, but the effect will
be limited to only a few e-foldings of the horizon crossing time.

\centerline{\bf Acknowledgements}

This work was partially supported by the European Union's Seventh 
Framework Programme (FP7-REGPOT-2012-2013-1) under grant agreement 
number 316165; by the European Union's Horizon 2020 Programme
under grant agreement 669288-SM-GRAV-ERC-2014-ADG;
by a travel grant from the University of Florida International Center, 
College of Liberal Arts and Sciences, Graduate School and Office of 
the Provost; by NSF grant PHY-1506513; and by the UF's Institute for 
Fundamental Theory. One of us (NCT) would like to thank AEI at the 
University of Bern for its hospitality while this work was partially 
completed.

\end{document}